\documentstyle[12pt]{article}
\def\hybrid{\topmargin 0pt      \oddsidemargin 0pt
        \parskip 5pt plus 1pt   \jot = 1.5ex}
\catcode`\@=11
\def\marginnote#1{}

\newcount\hour
\newcount\minute
\newtoks\amorpm
\hour=\time\divide\hour by60
\minute=\time{\multiply\hour by60 \global\advance\minute by-\hour}
\edef\standardtime{{\ifnum\hour<12 \global\amorpm={am}%
        \else\global\amorpm={pm}\advance\hour by-12 \fi
        \ifnum\hour=0 \hour=12 \fi
        \number\hour:\ifnum\minute<10 0\fi\number\minute\the\amorpm}}
\edef\militarytime{\number\hour:\ifnum\minute<10 0\fi\number\minute}

\def\draftlabel#1{{\@bsphack\if@filesw {\let\thepage\relax
   \xdef\@gtempa{\write\@auxout{\string
      \newlabel{#1}{{\@currentlabel}{\thepage}}}}}\@gtempa
   \if@nobreak \ifvmode\nobreak\fi\fi\fi\@esphack}
        \gdef\@eqnlabel{#1}}
\def\@eqnlabel{}
\def\@vacuum{}
\def\draftmarginnote#1{\marginpar{\raggedright\scriptsize\tt#1}}

\def\draft{\oddsidemargin -0.1truein
        \def\@oddfoot{\sl preliminary draft \hfil
        \rm\thepage\hfil\sl\today\quad\militarytime}
        \let\@evenfoot\@oddfoot \overfullrule 3pt
        \let\label=\draftlabel
        \let\marginnote=\draftmarginnote
   \def\@eqnnum{{\rm (\theequation)}\rlap{\kern\marginparsep\tt\@eqnlabel}%
\global\let\@eqnlabel\@vacuum}  }

\newdimen\linethick  \linethick=0.4pt
\newdimen\hboxitspace    \hboxitspace=5pt
\newdimen\vboxitspace    \vboxitspace=5pt

\def\fr#1{%
\beq\new
\vcenter{
\hrule height\linethick
           \hbox{\vrule width\linethick
                 \kern\hboxitspace
                 \vbox{\kern\vboxitspace
                       \hbox{$\begin{array}{c}\displaystyle#1
          \end{array}$}%
                       \kern\vboxitspace}%
                 \kern\hboxitspace
                 \vrule width\linethick}%
           \hrule height\linethick}%
\eeq}

\newdimen\Squaresize \Squaresize=14pt
\newdimen\Thickness \Thickness=0.5pt

\def\Square#1{\hbox{\vrule width \Thickness
   \vbox to \Squaresize{\hrule height \Thickness\vss
      \hbox to \Squaresize{\hss#1\hss}
   \vss\hrule height\Thickness}
\unskip\vrule width \Thickness}
\kern-\Thickness}

\def\Vsquare#1{\vbox{\Square{$#1$}}\kern-\Thickness}

\def\numberbysection{\@addtoreset{equation}{section}
        \def\theequation{\thesection.\arabic{equation}}}
\numberbysection

\renewcommand{\theequation}{\thesection.\arabic{equation}}
\newcommand{\l@qq}[2]{\addvspace{2em}
 \hbox to\textwidth{\hspace{1em}\bf #1 \dotfill #2}}

\newcounter{app}

\def\app{\setcounter{equation}{0}
\def\theequation{\Alph{app}.\arabic{equation}}\par
   \addvspace{4ex}
   \@afterindentfalse
  \secdef\@app\@dapp}
\newcommand\@app{\@startsection {app}{1}{0ex}%
                                   {-3.5ex \@plus -1ex \@minus -.2ex}%
                                   {2.3ex \@plus.2ex}%
                                   {\normalfont\Large\bf}}

\def\@dapp#1{%
{\parindent \z@ \raggedright  \bf #1}\par\nobreak}
\def\l@app#1#2{\ifnum \c@tocdepth >\z@
    \addpenalty\@secpenalty
    \addvspace{1.0em \@plus\p@}%
    \setlength\@tempdima{2.5em}%
    \begingroup
      \parindent \z@ \rightskip \@pnumwidth
      \parfillskip -\@pnumwidth
      \leavevmode \bfseries
      \advance\leftskip\@tempdima
      \hskip -\leftskip
      #1\nobreak\hfil \nobreak\hb@xt@\@pnumwidth{\hss #2}\par
    \endgroup\fi}
\newcounter{sapp}[app]

\def\sapp{\def\theequation{\Alph{app}.\arabic{equation}}\par
   \@afterindentfalse
  \secdef\@sapp\@dsapp}
\newcommand\@sapp{\@startsection{sapp}{2}{\z@}%
                                     {-3.25ex\@plus -1ex \@minus -.2ex}%
                                     {1.5ex \@plus .2ex}%
                                     {\normalfont\large\bfseries}}

\def\@dsapp#1{%
{\parindent \z@ \raggedright  \bf #1}\par\nobreak}
\newcommand{\l@sapp}{\@dottedtocline{2}{1.5em}{3em}}

\def\titlepage{\@restonecolfalse\if@twocolumn\@restonecoltrue\onecolumn
     \else \newpage \fi \thispagestyle{empty}\c@page\z@
        \def\thefootnote{\fnsymbol{footnote}} }

\def\endtitlepage{\if@restonecol\twocolumn \else  \fi
        \def\thefootnote{\arabic{footnote}}
        \setcounter{footnote}{0}}  
\relax

\hybrid

\newtoks\@stequation

\def\subequations{\refstepcounter{equation}%
  \edef\@savedequation{\the\c@equation}%
  \@stequation=\expandafter{\theequation}
  \edef\@savedtheequation{\the\@stequation}
  \edef\oldtheequation{\theequation}%
  \setcounter{equation}{0}%
  \def\theequation{\oldtheequation\alph{equation}}}

\def\endsubequations{%
  \setcounter{equation}{\@savedequation}%
  \@stequation=\expandafter{\@savedtheequation}%
  \edef\theequation{\the\@stequation}%
  \global\@ignoretrue}

\makeatletter
\newdimen\normalarrayskip              
\newdimen\minarrayskip                 
\normalarrayskip\baselineskip
\minarrayskip\jot
\newif\ifold             \oldtrue            \def\new{\oldfalse}
\def\arraymode{\ifold\relax\else\displaystyle\fi} 
\def\eqnumphantom{\phantom{(\theequation)}}     
\def\@arrayskip{\ifold\baselineskip\z@\lineskip\z@
     \else
     \baselineskip\minarrayskip\lineskip1\baselineskip\fi}

\def\@arrayclassz{\ifcase \@lastchclass \@acolampacol \or
\@ampacol \or \or \or \@addamp \or
   \@acolampacol \or \@firstampfalse \@acol \fi
\edef\@preamble{\@preamble
  \ifcase \@chnum
     \hfil$\relax\arraymode\@sharp$\hfil
     \or $\relax\arraymode\@sharp$\hfil
     \or \hfil$\relax\arraymode\@sharp$\fi}}

\def\@array[#1]#2{\setbox\@arstrutbox=\hbox{\vrule
     height\arraystretch \ht\strutbox
     depth\arraystretch \dp\strutbox
     width\z@}\@mkpream{#2}\edef\@preamble{\halign \noexpand\@halignto
\bgroup \tabskip\z@ \@arstrut \@preamble \tabskip\z@ \cr}%
\let\@startpbox\@@startpbox \let\@endpbox\@@endpbox
  \if #1t\vtop \else \if#1b\vbox \else \vcenter \fi\fi
  \bgroup \let\par\relax
  \let\@sharp##\let\protect\relax
  \@arrayskip\@preamble}
\def\eqnarray{\stepcounter{equation}%
              \let\@currentlabel=\theequation
              \global\@eqnswtrue
              \global\@eqcnt\z@
              \tabskip\@centering                      
              \let\\=\@eqncr
              $$%
            \halign to \displaywidth  \bgroup
             \eqnumphantom \@eqnsel
      \hskip\@centering                               
    $\displaystyle  \tabskip\z@ {##}$%
    &\global\@eqcnt\@ne \hskip 2\arraycolsep
         $ \displaystyle  \arraymode{##}$\hfil
    &\global\@eqcnt\tw@ \hskip 2\arraycolsep
         $\displaystyle\tabskip\z@{##}$\hfil
         \tabskip\@centering
    &{##}\tabskip\z@\cr}
\makeatother

\def\bea{\begin{eqnarray}}
\def\eea{\end{eqnarray}}

\def\beq{\begin{equation}}
\def\eeq{\end{equation}}
\def\be{\beq\new\begin{array}{c}}
\def\ee{\end{array}\eeq}
\def\bse{\begin{subequations}}                
\def\ese{\end{subequations}}                 %

\begin{document}
\hspace{10cm}{ITEP ??/??}\\
\vspace{0.5cm}
\begin{center}
{\LARGE \bf
A Candidate for Solvable Large $N$  } \\
\vspace{0.5cm}{\LARGE \bf Lattice Gauge Theory in $D\geq{2}$. } \\
\vspace{0.7cm} {\large\bf Andrey Yu. Dubin}\\
\vspace{0.5cm}{\bf ITEP, B.Cheremushkinskaya 25, Moscow 117259, Russia}\\
\vspace{0.5cm}{{\it e-mail: dubin@vxitep.itep.ru}}\\
{tel.:7 095 129 9674}\\
{fax:7 095 883 9601}
 \end{center}
\vspace{0.6cm}

\begin{abstract}

I propose a class of $D\geq{2}$ lattice {\it SU(N)} gauge theories dual to
certain vector models endowed with the $local$ $[U(N)]^{D}$
conjugation-invariance and ${\bf Z_{N}}$ gauge symmetry. In the latter models,
both the partitition function and Wilson loop observables depend nontrivially
only on the $eigenvalues$ of the link-variables. Therefore, the
vector-model facilitates a master-field representation of the 
large $N$ loop-averages in the corresponding induced gauge system.
As for the partitition function, in the limit
$N\rightarrow{\infty}$ it is reduced to the $2Dth$ power of an effective
$one$-matrix $eigenvalue$-model which makes the associated
phase structure accessible. In particular a simple scaling-condition, that
ensures the proper continuum limit of the induced gauge
theory, is proposed. We also derive a closed expression for
the large $N$ average of a generic nonself-intersecting Wilson loop
in the $D=2$ theory defined on an arbitrary $2d$ surface.

\end{abstract}

\begin{center}
\vspace{0.5cm}{Keywords: Lattice, Yang-Mills, Duality, Solvability,
Phase transitions}\\
\vspace{0.2cm}{PACS codes 11.15.Ha; 11.15.Pg}
\end{center}

\newpage

\section{Introduction}

The theory of confinement in the $D=4$ continuum $SU(3)$ Yang-Mills gauge
system endures as a tantalizing challenge over the last few decades. One
possible
strategy to deal with this problem was introduced in the seminal paper
\cite{Wils} of K.Wilson. The general idea is to consider a lattice
reformulation of the continuum $YM$ theory. In the right universality
class, the lattice system undergoes (for certain critical value $g_{0}^{2}$ of
the bare coupling constant $g^{2}$) a phase transition characterized by the
'divergence' of the correlation length. In this case the continuum $YM$
theory (supposed to be unique in $D=4$) can be recovered as the low-energy
theory at the scales much larger than the lattice spacing, with $g^{2}$
approaching $g_{0}^{2}$ along the $Wilsonian$ RG trajectory.

The employed so far $D>2$ lattice $YM$ theories are too
unwieldy to handle analytically. Consider for example
the most popular subvariety of the one-plaquette $SU(N)$ lattice
actions. The latter is defined by associating to each site ${\bf x}$ a factor
\be
\sum_{\{R_{\mu\nu}\}}~e^{-F(\{R_{\mu\nu}\})}~
\prod_{\mu\nu=1}^{D(D-1)/2} \chi_{R_{\mu\nu}}({U}_{\mu}({\bf x})
{U}_{\nu}({\bf x+\mu}){U}^{+}_{\mu}({\bf x+\nu})){U}^{+}_{\nu}({\bf x})),
\label{6.25m}
\ee
where the multiple Fourier expansion (in terms of $SU(N)$ irreducible
representations $R_{\mu\nu}$) involves the holonomies wrapped around
$D(D-1)/2$ elementary plaquettes. Besides the space-time dependence,
the models like (\ref{6.25m})
depend nontrivially on the $O(N^{2})$ degrees of freedom: not
only on the $O(N)$ eigenvalues $e^{i\omega_{j}(\rho)},~j=1,...N,$ but also on
the $nondiagonal$ ('angular') components $\Omega_{\rho}$ of the relevant
link-matrices $U_{\rho}=\Omega_{\rho}~diag[e^{i\omega(\rho)}]~
\Omega_{\rho}^{+},~\rho=1,...,D$.

So far our analytical knowledge about
systems of this type has been very limited, with the proper continuum limit of
the $D>2$ lattice gauge systems have been remaining beyond grasping. Besides
the $YM_{2}$ on a $2d$ surface \cite{Migd75,Rus,Witt91}, the
available solvable examples are mainly restricted to the situations
\cite{Konts,Kaz&Mig} where the model can be transformed into some
{\it eigenvalue-theory} of $q$ (hermitean or unitary) matrices.
The goal of the present paper is to take the reverse way around and 
$induce$ lattice gauge theories from vector-field models of the
eigenvalue-type. By definition, the (effective) action of the latter is
formulated in terms of such variables that makes manifest the invariance under
the set of $\alpha$-$dependent$ $[U(N)]^{\oplus q}$-conjugations
\be
S(\{\Psi_{\alpha}\tilde{W}_{\alpha}\Psi^{+}_{\alpha}\})=
S(\{\tilde{W}_{\alpha}\})=\sum_{\{R_{\alpha}\}} f(\{R_{\alpha}\})
\prod_{\alpha=1}^{q} \chi_{R_{\alpha}}(\tilde{W}_{\alpha})~,
\label{1.31}
\ee
where $\Psi_{\alpha},~\alpha=1,..,q,$ is the set of arbitrary $U(N)$ matrices.
In other words, the $largest$ possible
conjugation-invariance (\ref{1.31}) leads to the crucial
simplification: the action
$S_{eff}(\{\tilde{W}_{\alpha}\})\sim{O(N^{2})}$ 
depends nontrivially only on the $O(N)$ associated {\it eigenvalues}
$e^{i\theta_{j}(\alpha)}$ of $q$ (unitary or hermitean) matrices
$\tilde{W}_{\alpha},~\alpha=1,...,q$.

To generate a consistent mapping onto the gauge models, I propose
to start with the lattice systems defined in terms of the
$SU(N)$ link-variables $U_{\rho}({\bf z})$ and invariant
under the $local$ $[U(N)]^{\oplus D}$ conjugation-invariance
\be
U_{\rho}({\bf z})\rightarrow
{h^{+}_{\rho}({\bf z})~U_{\rho}({\bf z})~h_{\rho}({\bf z})}~~,~~
h_{\rho}\in{U(N)}~~,~~\rho=1,...,D,
\label{2.50}
\ee
combined with the reduced
gauge symmetry with respect to the center $T$ of the Lie group
\be
U_{\rho}({\bf z})\rightarrow
{H^{+}({\bf z})~U_{\rho}({\bf z})~H({\bf z+\rho})}~~,~~
H({\bf z})\in{T}~~,~~T={\bf Z_{N}}~,
\label{2.50b}
\ee
and, as a result of (\ref{2.50b}), with the global
$[{\bf Z_{N}}]^{D}$-invariance
\be
[T]^{\oplus D}:~U_{\rho}({\bf z})\rightarrow{t_{\rho}U_{\rho}({\bf z})}~~,~~
t_{\rho}\in{T={\bf Z_{N}}}~.
\label{2.9}
\ee
This class of eigenvalue-systems, being reduced to the 'one-plaquette'
subvariety, can be defined as following. Generalizing eq. (\ref{6.25m}), one
is to associate to each site ${\bf x}$ the factor
\be
\sum_{\{R_{\mu\nu}\}} e^{-S}
\prod_{\{\mu\nu\}} \chi_{R_{\mu\nu}}({U}_{\mu}({\bf x}))
\chi_{R_{\mu\nu}}({U}_{\nu}({\bf x+\mu}))
\chi_{R_{\mu\nu}}({U}^{+}_{\mu}({\bf x+\nu}))
\chi_{R_{\mu\nu}}({U}^{+}_{\nu}({\bf x})),
\label{6.25c}
\ee
(where $S\equiv{S(\{R_{\mu\nu}\})}$) in compliance with the pattern
(\ref{1.31}). What is even more important, owing to
(\ref{2.50}) the average of a (nonself-intersecting) Wilson loop
 \be
 W_{C}(U)=
 tr(U_{\mu}({\bf x})U_{\nu}({\bf x+\mu})... U_{\rho}({\bf x-\rho}))=
 tr(\prod_{\{{\bf z_{k}}\in{C}\}} U_{\rho_{k}}({\bf z_{k}})~)
 \label{2.13}
 \ee
can be also directly expressed, as we will see, in terms of the corresponding
eigenvalues
 \be
 <W_{C}(U)>=N^{1-L_{C}}~<\prod_{\{{\bf z_{k}}\in{C}\}}  tr(U_{\rho_{k}}({\bf z_{k}}))>~.
 \label{2.13b}
 \ee

The mapping of the model like (\ref{6.25c}) onto the corresponding gauge
system is to be performed in the two steps. First, we introduce
the auxiliary $SU(N)$ scalar field $\tilde{G}({\bf z})$ (assigned to the
lattice sites)
\be
U_{\rho}({\bf z})\rightarrow{
\tilde{G}^{+}({\bf z})~U_{\rho}({\bf z})~\tilde{G}({\bf z+\rho})}~~,~~
\rho=1,...,D,
\label{2.0z}
\ee
via the 'gauge-transformation' that leaves invariant both the $SU(N)$
measure $DU_{\rho}({\bf z})$ and the Wilson loop observables (\ref{2.13}).
Second, one integrates over the scalar
field $\tilde{G}({\bf z})$ with the Haar measure (normalized by
$\int d\tilde{G}({\bf z})=1$) that results in the associated effective theory
with the manifestly gauge-invariant action
$\tilde{S}_{eff}(\{U_{\rho}({\bf z})\})$. Indeed, after the extension
(\ref{2.0z}) the corresponding action
$S(\{\tilde{G}^{+}({\bf z})U_{\mu}({\bf z})\tilde{G}({\bf z+\mu})\})$
is invariant under the gauge transformations
\be
\bar{U}_{\mu}({\bf z})={g^{+}({\bf z})~U_{\mu}({\bf z})~g({\bf z+\mu})}~~,~~
\bar{G}({\bf z})={g^{+}({\bf z})~\tilde{G}({\bf z})}~,
\label{6.26}
\ee
formulated for $\tilde{G}({\bf z})$ in the unconventional way.
The crucial consequence of the 'extended' gauge symmetry (\ref{6.26}) is that
the induced $\tilde{S}_{eff}(\{U_{\rho}({\bf z})\})$
necessarily assumes the conventional form of the $multi$-plaquette lattice
$YM$ action composed from the generic $closed$ Wilson loops (\ref{2.13}).
In other words, the Haar integrations over $\tilde{G}({\bf z})$
$intertwine$ the contractions of the ${U}_{\rho}({\bf z})$-tensors to
recollect the latter into the gauge-invariant combinations (\ref{2.13}). 
For example, a single $\mu\nu$-component of the block-product in eq.
(\ref{6.25c}) after the four different $\tilde{G}({\bf z})$-integratioons (
refered to the sites of the corresponding elementary $\mu\nu$-plaquette) is
mapped onto
\be
\sim{\chi_{R_{\mu\nu}}({U}_{\mu}({\bf x}){U}_{\nu}({\bf x+\mu})
{U}^{+}_{\mu}({\bf x+\nu}){U}^{+}_{\nu}({\bf x}))}~.
\label{6.25y}
\ee

By construction, the gauge systems induced via (\ref{2.0z}) from the
eigenvalue-models (\ref{6.25c}) (invariant under (\ref{2.50})-(\ref{2.9})) are
endowed with the $same$ set of symmetries as the conventional lattice gauge
theories like the Wilson's one. In this perspective, the local
$[U(N)]^{\oplus D}$-symmetry (\ref{2.50}) can
be viewed as the 'hidden' symmetry inherent (via (\ref{2.0z})) in the proposed
family of the induced gauge theories. It is due to this symmetry the number of
the active degrees of freedom {\it per site} is thus substantialy reduced 
\be
O((D-1)N^{2})\rightarrow{O(DN)}
\label{6.25n}
\ee
as the action (\ref{6.25c}) does not
depend on the $nondiagonal$ components $\Omega_{\rho}({\bf z})$ of
$U_{\rho}({\bf z})=\Omega_{\rho}({\bf z})~diag[e^{i\theta(\rho|{\bf z})}]~
\Omega_{\rho}^{+}({\bf z}),~\rho=1,...,D$. Nevertheless, even after this
considerable reduction the model (\ref{6.25c}) is still not transparent
enough to allow for an exact solution. This forces us to search for
potentially solvable cousins of the $SU(3)$ eigenvalue-systems
among their $\lim_{N\rightarrow{\infty}} SU(N)$ counterparts keeping certain
analogue of the 't Hooft coupling $g^{2}N$ to be $\sim{O(N^{0})}$.
In this limit, due to (\ref{2.13b}) the eigenvalue-model (\ref{6.25c}) can be
viewed as a realization of the master-field representation for the large $N$
loop-averages (\ref{2.13}) in the associated (via (\ref{2.0z})) induced gauge
theory.

Indeed, in the large $N$
limit the $\Omega_{\rho}({\bf z})$-independence of (\ref{6.25c}) becomes
particularly advantageous. Let the weight $e^{-S(\{R_{\mu\nu}\})}$ in eq.
(\ref{6.25c}) be a generic function consistent with the $O(N^{2})$-scaling of
the free energy. Owing to the reduciton (\ref{6.25n}) of the active
degrees of freedom, in the computation of the large $N$ partitition function
and loop-averages (\ref{2.13b}) one can employ the good old
saddle-point method. The latter is to be applied $either$ to the $D\times
(N-1)$
eigenvalues $diag[e^{i\omega(\rho|{\bf z})}]$ of $U_{\rho}({\bf z})$
themselves (akin to
\cite{Gr&Witt}) $or$ to their Fourier duals (similarly to \cite{Dougl&Kaz}) -
the $SU(N)$ representations $R_{\mu\nu}({\bf z})$  parametrized by the
corresponding sets of $N-1$ integers
$\lambda_{j}(\mu\nu |{\bf z}),~j=1,...,N-1$. To generate additional
simplifications, we constrain the weight
$e^{-S(\{R_{\mu\nu}\})}$ to be invariant under the group-product
\be
S(D(D-1)/2)\otimes {\bf Z_{2}}~~~~;~~~~{\bf Z_{2}}:~ 
\otimes_{\{\mu\nu\}} R_{\mu\nu}\leftrightarrow{
\otimes_{\{\mu\nu\}} \bar{R}_{\mu\nu}}.
\label{6.3pc}
\ee
It combines the group of permutations within the $\{\mu\nu\}$-set 
of the irrep-indices with the $simultaneous$
${\bf Z_{2}}$-conjugation ($\chi_{R}(U^{+})=\chi_{\bar{R}}(U)$) of $all$ the
involved irreps. The ${\bf Z_{2}}$-symmetry plays, as we will see, an
important role for the consistency of the large $N$ construction.

Altogether, the partitition function $\tilde{X}_{L^{D}}$ of the large $N$
lattice $YM$ theory induced from (\ref{6.25c}) (in a $D$-volume $L^{D}$) 
is reproduced
\be                 
\lim_{N\rightarrow{\infty}} \tilde{X}_{L^{D}}=\lim_{N\rightarrow{\infty}}
(\tilde{X}_{r})^{L^{D}}~,
\label{2.10}
\ee
through the partitition function $X_{r}$ of the associated
$D(D-1)/2$-plaquette $SU(N)$ model with the reduced space-time dependence
\be
e^{-S_{r}(\{U_{\rho}\})}=\sum_{\{R_{\mu\nu}\}}e^{-S(\{R_{\mu\nu}\})}
\prod_{\{\mu\nu\}}
|\chi_{R_{\mu\nu}}(U_{\mu})\chi_{R_{\mu\nu}}(U_{\nu})|^{2}
\label{1.35}
\ee
which is invariant under $[{\bf Z_{N}}]^{D}$ symmetry
\be
[T]^{\oplus D}:~U_{\rho}\rightarrow{t_{\rho}U_{\rho}}~~,~~
t_{\rho}\in{T={\bf Z_{N}}}~,
\label{2.9x}
\ee
and formulated in terms of the eigenvalues of $D$ different link-variables
$U_{\rho}$ instead of the $DL^{D}$ original eigenvalues of
$U_{\rho}({\bf z})$. It is instructive to rewrite 
first the large $N$ partitition function $\tilde{X}_{r}$ as the specific
{\it generating functional} 
\be
\lim_{N\rightarrow{\infty}} \tilde{X}_{r}=\lim_{N\rightarrow{\infty}}
[~\sum_{\{R_{\rho}\}}\sum_{\{R_{\mu\nu}\}}
e^{-S(\{R_{\mu\nu}\})/2}
\otimes_{\rho=1}^{D} L^{(D-1)}_{R_{\rho}|\{R_{\rho\nu}\}}~]^{2}
\label{6.3p}
\ee
of the $D$-products of the generalized Littlewood-Richardson (GLR)
coefficients of the $(D-1)th$ order
\be
L^{(D-1)}_{R_{\rho}|\{R_{\rho\nu}\}}=
\int dU^{SU(N)}_{\rho}~\chi_{R_{\rho}}(U^{+}_{\rho})~
[\otimes_{\nu\neq{\rho}}^{D-1}\chi_{R_{\rho\nu}}(U_{\rho})]~\in{~\bf
Z_{\geq{0}}}
\label{6.3}
\ee
assuming nonnegative $integer$ values.

As we will see, the invariance of $e^{-S(\{R_{\mu\nu}\})}$ under
(\ref{6.3pc}) is sufficient for the reduction of the large $N$ representation
(\ref{6.3p}) to the $2Dth$ power
\be
\lim_{N\rightarrow{\infty}} \tilde{X}_{r}=\lim_{N\rightarrow{\infty}}
[~{\sum_{R(\{\lambda\})}}''~e^{-S(R|D)}~]^{2D}
\label{1.2b}
\ee
of an effective $one$-matrix $SU(N)$ model formulated in terms of the
$\{\lambda\}$-set of $N-1$ integers $\lambda_{j},~j=1,...,N-1,$ which
parametrize canonically the $SU(N)$ irreps $R\equiv{R(\{\lambda\})}$. The sum
${\sum_{R}}''$,
being the large $N$ 'image' of the $D$ summations over $\{R_{\rho}\}$ in eq.
(\ref{6.3p}), is dynamically constrained by the following condition.
Let $n(R)$ stand for the number of boxes in the $SU(N)$ Young tableau
$Y^{(N)}_{n(R)}$ associated to $R$. Then, in (\ref{1.2b})
both $n(R)$ and $n(\bar{R})$ must be nonnegative multiples of $(D-1)$.
Next, taking advantage of the freedom to choose
$S(\{R_{\mu\nu}\})$ in eq. (\ref{6.3p}), the
${\bf Z_{2}}$-invariant effective action
$S(R|D)=S(\bar{R}|D)$ in (\ref{1.2b}) can be judiciously selected in $any$
required form consistent with $-ln[\tilde{X}_{r}]\sim{O(N^{2})}$. In what
follows, our attention will be restricted to the simplest $solvable$ class of
the $SU(N)$ or $U(N)$ models with $S(R|D)$ being defined as
\be
e^{-S(\{\lambda\})}=|dimR(\{\lambda\})|^{q}~
e^{-\sum\limits_{n=1}^{M_{0}}\sum\limits_{\tilde{r}\in{Y_{2n}}}
g_{\tilde{r}(\{p\})}
\prod\limits_{k=1}^{2n}
[\sum\limits_{i=1}^{N}(\lambda_{i}-\frac{N-1}{2})^{k}]^{p_{k}}}~,
\label{4.1}
\ee
where $q>0$. In the $SU(N)$ case the set $\{g_{\tilde{r}(\{p\})}\}$ is
supposed to be invariant under the translations $\lambda_{i}\rightarrow
{\lambda_{i}+m}$ to match with the standard $SU(N)$ condition
$\lambda_{N}=0$. As for
the sum $\sum_{\tilde{r}}$ in the exponent, it runs over the irreps
$\tilde{r}\equiv{\tilde{r}(\{p\})}\in{Y_{2n}}$ of the even symmetric group
$S(2n)$
(labelled by the partititions $\{p\}=[1^{p_{1}}2^{p_{2}}...2n^{p_{2n}}]$ of
$2n$: $\sum_{k=1}^{2n} kp_{k}=2n$) with $n\leq{M_{0}}\in{{\bf Z_{\geq{1}}}}$.

Finally, the organization of the paper is as following. In Section 2 we
present the details of the large $N$ reduction (\ref{2.10}) representing 
$X_{r}$ in the form of the GLR generating functional (\ref{6.3p}).
In Section 3, the transformation of the large $N$ GLR functional (\ref{6.3p})
into the 1-matrix representation (\ref{1.2b}) is explicitly performed, and
the large $N$ scaling properties of the selected 1-matrix family (\ref{1.2b})
are formulated. The continuum limit (CL) in the gauge theories induced from
the models like (\ref{6.25c}) is analysed in Section 4. We
propose a simple criterion for the reduced 1-matrix system (\ref{4.1})
to ensure the $localization$ $\{U_{\rho}\rightarrow{\hat{1}}\}$ (modulo
(\ref{2.9x})) in the infinitesimal
vicinity, scaling as $O(N^{0})$, of the group-unity $\hat{1}$. In turn, it
predetermines that in the
associated induced gauge theory the link-variables are as well localized
$\{U_{\rho}({\bf z})\rightarrow{\hat{1}}\}$ (modulo (\ref{2.9}) and the
gauge symmetry) which is tantamount to the proper CL.
The phenomenon of the auxiliary 'continuum' limits (which accompany the
conventional CL) in the effective system (\ref{1.2b}) is also
discussed in connection with the large $N$ phase transitions (PT).

In Section 5 we derive a closed expressions for the large $N$ average of a
generic nonself-intersecting Wilson loop in the two-dimensional
eigenvalue-system (\ref{6.25c}) on an arbitrary $2d$ discretized closed
surface. Building on this expression, we associate to a given 
eigenvalue-system (\ref{6.25c}) the corresponding gauge theory (\ref{6.25m})
in such a way that both of them are supposed to have (in $D=2$) the same CL.
The general pattern of averages (including self-intersecting loops and
irreducible multi-loop correlators) is discussed in Section 6.
Some peculiarities of the large $N$ pattern of the (multi)loop averages
(\ref{2.13})
are revealed and interpreted. In the last section we make our conclusions and
discuss possible directions of further research. The Appendices
contain
technical details employed in the main text.

\section{Large $N$ reduction of the GLR Functional.}
\setcounter{equation}{0}

Let us proceed with the derivation of the reduced representation (\ref{2.10})
of the large $N$ partitition function $\tilde{X}_{L^{D}}$ of the proposed
eigenvalue-model
(\ref{6.25c}). We will demonstrate also that reduced actions like
(\ref{1.35}), in fact, reproduce partitition function of a {\it larger} family
of eigenvalue-systems including the 'multi-plaquette' generalizations of
(\ref{6.25c}).

To begin with, one notes that in the theory (\ref{6.25c}), $X_{L^{D}}$ can be
readily represented in the form which associates to each site ${\bf z}$ of
the base-lattice the properly weighted sum of the blocks
\be
\otimes_{\rho=1}^{D}\int dU_{\rho}~\otimes_{\nu\neq{\rho}}^{D-1}
[~\chi_{R_{\rho\nu}({\bf z-\nu})}(U^{+}_{\rho}({\bf z}))~
\chi_{R_{\rho\nu}({\bf z})}(U_{\rho}({\bf z}))~]
\label{6.3n}
\ee
composed of the characters which in (\ref{6.25c}) are refered to
the plaquettes sharing the $D$ $\rho$-links based at ${\bf z}$.
Employing the standard fusion-rules of the Lie group characters
\be
\otimes_{k=1}^{p}~\chi_{R_{k}}(V)=\sum_{k=1}^{p}~\chi_{R_{+}}(V)~
L^{(p)}_{R_{+}|\{R_{k}\}}~,
\label{6.3a}
\ee
the combination (\ref{6.3n}) can be rewritten in terms of 
the GLR coefficients
\be
\otimes_{\rho=1}^{D}~
L^{(D-1)}_{R_{\rho}({\bf z})|\{R_{\rho\nu}({\bf z-\nu})\}}~
L^{(D-1)}_{R_{\rho}({\bf z})|\{R_{\rho\nu}({\bf z})\}}~.
\label{6.3m}
\ee
Consider the representation of $\tilde{X}_{L^{D}}$ in terms of the GLR blocks
(\ref{6.3m}) and let the number of sites $L^{D}<<N$ while
$-ln[\tilde{X}_{L^{D}}]\sim{O(N^{2})}$.
The remaining summations over $R_{\mu\nu}({\bf z}),R_{\rho}({\bf z})$, being
parametrized by the $O(L^{D}N)$ integers, in the large $N$ limit can be
evaluated with the help of the
saddle-point (SP) approach (in a manner akin to \cite{Dougl&Kaz}; see also
Section 4). Since the SP irreps are supposed to be space-time independent
\be
R^{(0)}_{\mu\nu}({\bf z})=R^{(0)}_{\mu\nu}~~,~~
R^{(0)}_{\rho}({\bf z})=R^{(0)}_{\rho}~,
\label{6.71}
\ee
the SP equations in the $L^{D}$-lattice theory (\ref{6.25c}) are reduced
to the SP equations derived from the functional (\ref{6.3p}) associated to
the reduced model (\ref{1.35}). In relating (\ref{6.3p}) and (\ref{1.35})
we have used also the ($\gamma=2$ case of the) identity
\be
\lim_{N\rightarrow{\infty}}
\int \prod_{\alpha=1}^{p} \sum_{R_{\alpha}}~
e^{-S_{p}(\{R_{\beta}\})}=
\lim_{N\rightarrow{\infty}}
[~\int \prod_{\alpha=1}^{p} \sum_{R_{\alpha}}~
e^{-S_{p}(\{R_{\beta}\})/\gamma}~]^{\gamma}~,
\label{3.45}
\ee
valid provided that $\gamma>0$, the saddle-point values of $both$
$e^{-S_{p}(\{R^{(0)}_{\beta}\})}$ $and$
$e^{-S_{p}(\{R^{(0)}_{\beta}\})/\gamma}$ are unique and
$positive$, while $-ln[\tilde{X}_{{p}}]\sim{O(N^{2})}$. Indeed, in this case the
equivalence of the saddle-point equations, associated to both sides of
(\ref{3.45}), is sufficient to prove this identity.

It is instructive to rederive the
$[\tilde{X}_{r}]^{L^{D}}\leftrightarrow{\tilde{X}_{L^{D}}}$ correspondence
(\ref{2.10}) applying the SP method directly to the eigenvalues
$diag[e^{i\omega(\rho|{\bf z})}]$ of $U_{\rho}({\bf z})$. The effective action
$S_{eff}(\{\omega_{j}(\rho|{\bf z})\})\sim{O(N^{2})}$, where by definition
\be
\tilde{X}_{L^{D}}=\int \prod_{j,\rho,{\bf z}}\frac{d\omega_{j}(\rho|{\bf z})}{2\pi}~
exp[-S_{eff}(\{\omega_{j}(\rho|{\bf z})\})]~,
\label{1.35b}
\ee
is composed of the two parts $S_{eff}=S_{0}+S_{J}$. The
'bare' part $S_{0}(\{\omega\})$ is associated to the factor (\ref{6.25c})
rewritten as $e^{-S_{0}(\{\omega\})}$. The remaining
contribution $S_{J}=-\sum_{\rho,{\bf z}}
ln[\Delta(\{\omega_{p}(\rho|{\bf z})\})]$ is due to
the Jacobian induced by the change of the variables
\be
\int dU=\int d\Omega~\prod_{k=1}^{N}\int_{-\pi}^{+\pi}
\frac{d\omega_{k}}{2\pi}~\Delta(\{\omega_{p}\})\equiv{\int
d\Omega J(T)dT}~,
\label{1.2d}
\ee
\be
\Delta(\{\omega_{p}\})=\delta^{[2\pi]}(\sum_{k=1}^{N}\omega_{k})
\prod_{i<j} |2sin(\frac{\omega_{i}-\omega_{j}}{2})|^{2}
~~,~~T=diag[e^{i\omega}],
\label{1.2m}
\ee
introducing, $U=\Omega~diag[e^{i\omega}]~\Omega^{+}$, the 'angular'
(nondiagonal) $\Omega$- and 'radial' (diagonal)
$diag[e^{i\omega}]$-components. According to
(\ref{1.2m}) and the Weyl character formula \cite{Dr&Zub}
\be
\chi_{R(\{\lambda\})}(U)=
det_{k,j}(e^{i\lambda_{k}\omega_{j}})/det_{k,j}(e^{i(N-k)\omega_{j}}),
\label{4.3}
\ee
the action $S_{eff}(\{\omega_{j}(\rho|{\bf z})\})$ (in addition to the
symmetries (\ref{2.50})-(\ref{2.9})) is invariant under the 
$S(N)$ Weyl group of the the permutations
\be
\omega_{j}(\rho|{\bf z})\rightarrow{\omega_{\sigma(j)}(\rho|{\bf z})}~~,~~
\sigma\in{S(N)}~.
\label{1.2x}
\ee
In what follows, the latter symmetry is supposed to be 'fixed' by the
Faddeev-Popov condition $\omega^{(0)}_{j}(\rho|{\bf z})>
\omega^{(0)}_{j+1}(\rho|{\bf z})$. Having imposed this constraint, there
still remains the whole $orbit$ of the SP solutions
$\omega^{(0)}_{j}(\rho|{\bf z})$ generated apparently by the finite $N$
symmetries (\ref{2.50b}),(\ref{2.9}).

The subtlety is that the leading
$O(N^{2})$ order of $S_{eff}(\{\omega^{(0)}_{j}(\rho|{\bf z})\})$ (restricted to the
to the SP values $\{\omega^{(0)}\}$) is invariant
under the extended $local$ $[{\bf Z_{N}}]^{\oplus D}$ symmetry
\be
[{\bf Z_{N}}]^{\oplus D}:~U_{\rho}({\bf z})\rightarrow
{t_{\rho}({\bf z})U_{\rho}({\bf z})}~~,~~
t_{\rho}({\bf z})\in{{\bf Z_{N}}}~,
\label{2.50m}
\ee
which is broken by the $subleading$ orders of $S_{eff}$ down to
(\ref{2.50b}),(\ref{2.9}).
We refer to Appendix A for the details and now simply make use of
(\ref{2.50m}). In particular, it implies that (having fixed (\ref{1.2x})) the
residual $SU(N)$ SP orbit
is supposed to be generated from a unique and space-time independent solution
$\omega^{(0)}_{j}(\rho|{\bf z})$ by the set of transformations
(\ref{2.50m}) which is larger than (\ref{2.50b}),(\ref{2.9}).
Employing for (\ref{1.35b}) the analogue of
the identity (\ref{3.45}) (readily modified to account for the degeneracy
(\ref{2.50m})), one arrives at the required correspondence (\ref{2.10})
between $[\tilde{X}_{r}]^{L^{D}}$ and ${\tilde{X}_{L^{D}}}$.

As a side remark, in the computation of a large $N$ average 
of a nonself-intersecting Wilson loop (\ref{2.13}) the subleading orders of
$S_{eff}(\{\omega_{j}(\rho|{\bf z})\})$ must be
necessarily included. It will be clear from Sections 5 and 6 (see also
Appendix A). Consequently, as far as the
$\lim_{N\rightarrow{\infty}}<W_{C}(U)>$ averages
is concerned, the extended symmetry (\ref{2.50m}) is $not$ observable
directly (although leading to certain peculiarities discussed in
Section 6).

Now we are ready to formulate the following correspondence that will be used
in Section 4 for the analysis of the continuum limit. The latter limit
takes place if in the (induced) gauge theory the link-variables are localized
in a vicinity, scaling as $O(N^{0})$, of the group-unity $\hat{1}$:
$U_{\rho}({\bf z})\rightarrow{\hat{1}}$ modulo (\ref{2.9}) and the $SU(N)$
gauge symmetry. Let in the associated reduced system (\ref{1.35}) the coupling
constants of $S_{r}(\{U_{\rho}\})$ are adjusted to provide with similar
localization $U_{\rho}\rightarrow{\hat{1}}$ (modulo (\ref{2.9x})).
The previous SP analysis implies that this constraint on
$S_{r}(\{U_{\rho}\})$ ensures that in the corresponding eigenvalue-system
(\ref{6.25c}) the link-variables $U_{\rho}({\bf z})$ (entering the Wilson
loop observables) are as well localized
\be
U_{\rho}\rightarrow{\hat{1}}~~
\bmod (\ref{2.9x})~\Longrightarrow{~U_{\rho}({\bf z})
\rightarrow{\hat{1}}~~
\bmod \{ (\ref{2.50b}),(\ref{2.9})\}}~,~
\forall{\rho}.
\label{6.70}
\ee
In turn, the pattern of the mapping (\ref{2.0z}) guarantees that in the
associated induced gauge theory the condition (\ref{6.70}) results in the
required localization of $U_{\rho}({\bf z})$.

Finally, let us note that the SP analysis (applied to
$diag[e^{i\omega(\rho|{\bf z})}]$) reveals the following direction to
generalize the 1-plaquette family (\ref{6.25c}) of the eigenvalue-models with
the partitition function reproduced via the reduced system (\ref{1.35}).
First, with the help of the Frobenius formula \cite{Dr&Zub} one is to expand
the characters entering (\ref{6.25c}) in terms of the trace products
$\prod_{k=1}^{n} [tr(U^{k}({\bf z}))]^{p_{k}}$. Consider a set of such
products associated to the plaquette-factor based at a particular site
${\bf x}$. The idea is to change the arguments in each $tr(U^{k}({\bf x}))$
separately trading ${\bf x}$ for generic sites. The
only natural restriction for such interchange is that the new set of the
trace products remains invariant under the $[{\bf Z_{N}}]$ gauge symmetry
(\ref{2.50b}). In this way, one constructs an eigenvalue-counterpart of the
gauge theories with the action depending on the multi-plaquette Wilson loops.
Provided the interchanges are performed in the same way for
all lattice reference-sites ${\bf z}$, the SP values of
$diag[e^{i\omega(\rho|{\bf z})}]$ are supposed to be translationally
invariant (modulo (\ref{1.2x}) and (\ref{2.50m})).
Consequently, by construction of the interchange the partitition function
of the deformed eigenvalue-system is still reproduced in the large $N$ limit
by the same reduced model (\ref{1.35}). Actually, there are many other
ways to generalize (\ref{6.25c}) preserving the structure of the associated
reduced system.

\section{The effective $N\rightarrow{\infty}$ 1-matrix theory.}
\setcounter{equation}{0}

Further reduction of the GLR functional (\ref{6.3p}) to the '1-matrix'
representation (\ref{1.2b}) is built on the localization of
the large $N$ summations over $\{R(\rho)\}\otimes\{R(\mu\nu)\}$ on the
solution
$\{R^{(0)}(\rho)\}\otimes\{R^{(0)}(\mu\nu)\}$ of
the corresponding saddle-point equations. We defer the discussion of these
equations untill the next section, and now simply assert the expected
properties of the solution in the case when the constraints (\ref{6.3pc}) are
additionally imposed. To be more specific, we select the option when
the effective 1-matrix system in eq. (\ref{1.2b}) is reduced to the family
(\ref{4.1}). Altogether, the
saddle-point $SU(N)$-set $\{R^{(0)}(\rho)\}\otimes\{R^{(0)}(\mu\nu)\}$,
\be
R^{(0)}(\rho)=R^{(0)}=\bar{R}^{(0)}~,~\forall{\rho}~~~;~~~
R^{(0)}(\mu\nu)=R_{2}^{(0)}=\bar{R}^{(0)}_{2}~,~\forall{\mu\nu},
\label{3.43b}
\ee
is supposed to be $unique$, $\{\rho\}\otimes\{\mu\nu\}$ independent
respectively, and {\it selfdual}. As a result, the generating functional
(\ref{6.3p}) is equivalent in the large $N$ limit to the $reduced$ system
obtained after the identification
\be
R(\mu\nu)\equiv{R_{2}}\in{Y_{n_{2}}^{(N)}}~~,~~
R(\rho)\equiv{R}\in{Y_{n}^{(N)}}~,
\label{3.43}
\ee
with the remaining summations over $R,~R_{2}$ being localized on the same
saddle-point values (\ref{3.43b}).

Since the reduced action
(\ref{1.35}) is composed of the blocks containing equal amounts of
$U_{\rho}$ and $U^{+}_{\rho}$ $SU(N)$ factors, $S_{r}(\{U_{\rho}\})$ is
invariant under the larger set of transformations (\ref{2.9x}) where
${\bf Z_{N}}$ is extended to $U(1)$. As a result, the partitition
function $\tilde{X}_{r}$ is invariant under the substitution of the $SU(N)$
link-variables by the $U(N)=[SU(N)\otimes U(1)]/{\bf Z_{N}}$ ones
\be
U^{SU(N)}_{\rho}\rightarrow{U^{U(N)}_{\rho}}~~,~~\forall{\rho}~.
\label{3.43m}
\ee
Complementary, the involved $SU(N)$ irreps $R_{\phi}$ can be viewed as
belonging to the (anti)chiral subset of the $U(N)$ ones.
Owing to (\ref{3.43m}), the sum in (\ref{6.3p}) over $SU(N)$ irreps $R$
is effectively constrained by the
${\bf Z_{2}}$-invariant pair of the $U(N)$ conditions (nontrivial in $D>2$)
\be
L^{(D-1)}_{R|R^{\oplus D-1}_{2}}\neq{0}~\Rightarrow{~n(R)=n(R_{2})(D-1)}
~~,~~n(\bar{R})=n(\bar{R}_{2})(D-1),
\label{3.46c}
\ee
predetermined by the pattern of $U(N)$ GLR coefficients corresponding to
(\ref{6.3}). In eq. (\ref{3.46c}), the integers $n(R_{\phi}),
n(\bar{R}_{\phi})\in{\bf Z_{\geq{0}}}$ denote the
number of boxes in the ${\bf Z_{2}}$-invariant pair of the Young tableaus
corresponding to (\ref{3.43}):
\be
n(R(\{\lambda\}))=\sum_{i=1}^{N} n_{i}=
\sum_{i=1}^{N}(\lambda_{i}-N+i).
\label{3.46m}
\ee
Recall also that the
irreps of $U(N)$ are labelled by a set of $N$ integers
$\lambda^{U(N)}_{i}$ (constrained by
$\sum_{i=1}^{N-1} \lambda_{i}=\lambda_{N} \bmod N$)
\be
\{\lambda^{U(N)}\}=\{\lambda_{1}+\lambda_{N}>...>
\lambda_{N-1}+\lambda_{N}>\lambda_{N}\}\in{[{\bf Z}^{\oplus N}/S(N)]}
\label{2.3}
\ee
generated from the $SU(N)$-sets of $N-1$ $nonnegative$ integers
$\{\lambda^{SU(N)}\}=\{\lambda_{1}>\lambda_{2}>..>\lambda_{N-1}>0\}$
by the extra integer number $\lambda_{N}\geq{0}$ or $\lambda_{N}<0$.
As for the ${\bf Z_{2}}$-conjugation $R\leftrightarrow{\bar{R}}$, it
reads
\be
\{\lambda_{i}\}\leftrightarrow
{\{-\lambda_{N-i+1}+\beta\}}~,
\label{4.6}
\ee
where $\beta^{U(N)}=(N-1)$ and $\beta^{SU(N)}=\lambda_{1}$ in the $U(N)$
and $SU(N)$ cases respectively.

Next, the effective action for $R,R_{2}$, resulting from the reduction
(\ref{3.43}), contains (according to (\ref{6.3p})) the $Dth$ power of
$L^{(D-1)}_{R|R^{\oplus D-1}_{2}}$. To simplify this
expression further, one can employ the ($\gamma=D$ case of the) identity
(\ref{3.45}). Altogether, it defines the $one$-matrix representation of the
large $N$ family (\ref{6.3p})
\be
\lim_{N\rightarrow{\infty}} \tilde{X}_{r}=\lim_{N\rightarrow{\infty}}
[~\int dU^{U(N)}\sum_{R,R_{2}} e^{-\frac{S^{(D)}(R_{2})}{2D}}
\chi_{R}(U^{+})~[\chi_{R_{2}}(U)]^{D-1}~]^{2D},
\label{3.44}
\ee
where the weight $S^{(D)}(R_{2})=S(\{R(\mu\nu)\})|_{\{R(\mu\nu)=R_{2}\}}$ is
deduced from that of (\ref{6.3p}) through the 'dimensional'
reduction (\ref{3.43}). The integrated in (\ref{3.44}) expression is
simply related to the associated reduced action $S_{r}(\{U_{\rho}\})$.
Consider first the case of the $separable$ weight $S(\{R(\mu\nu)\})=
\sum_{\{\mu\nu\}}S(R(\mu\nu))$ which
results in $S^{(D)}(R_{2})=D(D-1)S(R_{2})/2$. Once the
${\bf Z_{2}}$-selfduality
(\ref{3.43b}) of $R_{2}^{(0)}$ takes place, the identity (\ref{3.45}) can be
applied once more (this time with $\gamma=(D-1)/4$)
\be
\sum_{R_{2}}[~e^{-S(R_{2})}
\chi^{4}_{{R}_{2}}(U)~]^{\frac{D-1}{4}}\rightarrow
{[~\sum_{R_{2}}e^{-S(R_{2})}|\chi_{{R}_{2}}(U)|^{4}~]
^{\frac{(D-1)}{4}}}=e^{-\frac{S_{r}(\{U\})}{2D}},
\label{3.46b}
\ee
where $e^{-S_{r}(\{U\})}\equiv
{e^{-S_{r}(\{U_{\rho}\})}}|_{\{U_{\rho}=U\}}$ stands for
the original reduced action (\ref{1.35}) with coinciding arguments. As a
result, eq. (\ref{3.44}) assumes the concise form
\be
\lim_{N\rightarrow{\infty}} \tilde{X}_{r}=
[~\sum_{R}~\int dU exp[{-\frac{S_{r}(\{U\})}{2D}}]~
\chi_{R}(U^{+})~]^{2D}~.
\label{3.46}
\ee
Upon a reflection, the representation (\ref{3.46}) remains valid (provided
the $S(D(D-1)/2)$-invariance (\ref{6.3pc}) of $S(\{R(\mu\nu)\})$)
for a generic, $not$ necessarily separable form of the weight
$S(\{R(\mu\nu)\})$.

It is appropriate to remark that the consistency of a substitution like
(\ref{3.46b}) requires the positivity
$\chi_{{R}^{(0)}_{2}}(diag[e^{i\omega^{(0)}}])=
[\chi_{{R}^{(0)}_{2}}(diag[e^{i\omega^{(0)}}])]^{+}>0$, where $R^{(0)}_{2}$
is defined by eq. (\ref{3.43b}) while $\omega^{(0)}_{i}$ is the saddle-point
$orbit$ on which the integral (\ref{3.44}) over the $U$-eigenvalues
$\omega_{i}$ is localized. Also, the solution $\omega^{(0)}_{i}$ is supposed
to be unique modulo the symmetries leaving $L^{(D-1)}_{R|R^{\oplus D-1}_{2}}$
invariant, i.e. the $S(N)$ Weyl group (\ref{1.2x}) combined with
the center-transformations (\ref{2.9x}).

Next, let us introduce the Fourier-dual partner $e^{-H(R|D)}$ of
$e^{-\frac{S_{r}(\{U\})}{2D}}$,
\be
e^{-H(R|D)}=\sum_{R_{2}} exp[{-\frac{S^{(D)}(R_{2})}{2D}}]~
L^{(D-1)}_{R|R^{\oplus D-1}_{2}}~,
\label{3.48}
\ee
so that $e^{-S_{r}(\{U\})/2D}=\sum_{R_{1}}~
e^{-H(R_{1}|D)}~\chi_{R_{1}}(U)$. As a result, employing the standard
orthogonality of the characters, eq. (\ref{3.46}) is readily transformed into
the required form (\ref{1.2b}). Note also that we have introduced an implicit
dependence of $H(R|D)$ on $D$.

Let ${\bf Z_{2}}$-invariant $e^{-{S^{(D)}(R_{2})}/{2D}}$-factor be an
arbitrary function of $R_{2}$ consistent with
$-ln[\tilde{X}_{r}]\sim{O(N^{2})}$.
The central question at this step is what are the general constraints
defining the whole $e^{-H(R|D)}$-family of weights induced from the
$e^{-{S^{(D)}(R_{2})}/{2D}}$-variety via (\ref{3.48}).
Upon a reflection, the pattern of the $U(N)$ GLR coefficients
ensures that the $only$ defining property of the $e^{-H(R|D)}$-family is
the ${\bf Z_{2}}$-invariant pair of the $D>2$ conditions
(\ref{3.46c}) on the involved $SU(N)$ irreps. In other words, given $any$
irrep $R\in{Y^{(N)}_{(D-1)m}}$ we
expect that there exists at least one irrep $R_{2}\in{Y^{(N)}_{m}}$ so that
$L^{(D-1)}_{R|R^{\oplus D-1}_{2}}\neq{0}$. It suggests to factorize the
constraints (\ref{3.46c}) out
\be
e^{-H(R|D)}=e^{-S(R|D)}~\sum_{k,\bar{k}\in{\bf Z}}
\delta_{n(R),[D-1]k}~\delta_{n(\bar{R}),[D-1]\bar{k}}~.
\label{3.49z}
\ee
Summarizing, for a $fixed$ $D$ a generic residual $R$-valued function 
$e^{-S(R|D)}$ (consistent with ${\bf Z_{2}}$-invariance $S(R|D)=
S(\bar{R}|D)$ and with the scaling $-ln[\tilde{X}_{r}]\sim{O(N^{2})}$) can be
induced through (\ref{3.48}) provided the judicious adjustment of
$e^{-S^{(D)}(R_{2})/2D}$.

To make the representation (\ref{3.49z})
practical, the two $periodic$ $Kronecker$ delta-functions are introduced via
'$\varepsilon$-regularization' of the Poisson resummation formula
\be
\sum_{k\in{\bf Z}} \delta_{n,[D-1]k}=\lim_{\varepsilon\rightarrow{0}}
\sum_{p\in{\bf Z}} exp[-\frac{f(n,p,\varepsilon)}
{(D-1)}]~\sqrt{\varepsilon/[D-1]}~~,~~
\label{3.49b}
\ee
\be
f(n,p,\varepsilon)=2\pi [\varepsilon p^{2}-i~np]~.
\label{3.49m}
\ee
Altogether, it results in the 1-matrix representation of the
$SU(N)$ $GLR$ functional (\ref{6.3p})
\be
\lim_{N\rightarrow{\infty}} \tilde{X}_{r}=\lim_{N\rightarrow{\infty}}
\lim_{\varepsilon\rightarrow{0}}~
[\frac{\varepsilon}{[D-1]}\sum_{p,\bar{p}\in{\bf Z}}~
\sum_{n\in{\bf Z_{\geq{0}}}}\sum_{R\in{Y_{n}^{(N)}}}~exp(-A)~]^{2D}
\label{3.49}
\ee
\be
A=S(R|D)+
\frac{f(n(R),p,\varepsilon)+f(n(\bar{R}),\bar{p},\varepsilon)}{D-1}.
\label{3.49t}
\ee
Remark that if the additional $D>2$ constrains
(\ref{3.46c}) in (\ref{1.2b}) were omitted, the $D\geq{2}$ pattern
(\ref{3.49}) of $\tilde{X}_{r}$ would be essentially two-dimensional.
Indeed, eq. (\ref{3.49}) would reduce to the $[2D]th$-power of the
partitition function associated (when in eq. (\ref{4.1}) $q=2$) to the
continuum $generalized$ $2d$ YM on a sphere \cite{Witt91,Rus}.

In conclusion, we note that similar analysis of the $U(N)$ reduced model
(\ref{1.35}) (with the sum running over the $U(N)$ irreps) results in the
$U(N)$ counterparts of eqs. (\ref{3.46}) and (\ref{3.49}).

\subsection{The large $N$ scaling in the selected family of the
1-matrix models.}

Let us now discuss what conditions make the large $N$ limit of the selected
1-matrix variety (\ref{4.1}) well defined and consistent with the scaling
$-ln[\tilde{X}_{r}]\sim{O(N^{2})},~|\lambda_{j}|\sim{O(N)}$. As we will see
in Section 4, the latter scaling is necessary to ensure the applicability of
the saddle-point method to the summations over the integer-valued
$\{\lambda\}$-fields parametrizing relevant irreps.

To begin with, for each $r(\{p\})\in{Y_{2n}}$ one is to perform the proper
change of the variables
\be
\bar{\lambda}_{j}=\lambda_{j}/N~~,~~
g_{r(\{p\})}=b_{r(\{p\})}N^{\gamma_{r}}~~,~~\gamma_{r(\{p\})}=
2-2n-\sum_{k=1}^{2n} p_{k}~,
\label{4.1b}
\ee
postulating that $b_{r(\{p\})}\sim{O(N^{0})}$.
Next, the Weyl character formula yields for the irrep dimension
\be
dimR(\{\lambda\})=\prod_{1\leq{i}<j\leq{N}} (\lambda_{i}-\lambda_{j})/(j-i),
\label{4.3b}
\ee
which foreshadows the constraint $\lambda_{i}>\lambda_{i+1}$ inherent in the
pattern (\ref{2.3}) of the irrep-parametrization. Combining the latter
constraint with
the adjustment (\ref{4.1b}), one concludes that the $O(N^{2})$-order of
$-ln(\tilde{X}_{r})$ is $necessarily$ accumulated by those of
$|\lambda_{j}|$ which are $\sim{O(N)}$:
\be
\{|\lambda_{i}|\sim{O(N)}\}~~~
\Longleftrightarrow{~~~S(\{\lambda\})\sim{O(N^{2})}}~,
\label{4.22}
\ee
which altogether justifies that the characteristic values of
$\bar{\lambda}_{j}=\lambda_{j}/N$ are $\sim{O(N^{0})}$. For definiteness,
in the derivation of eq. (\ref{4.22}) we have supposed that
$S(\{\lambda\})|_{\{\lambda_{j}=({N-1})/{2}\}}\sim{O(N^{0})}$ when
$\lambda_{j}$ assume (unadmissible) coinciding values
$\lambda_{j}=({N-1})/{2},~\forall{j}$. This choice of the additive
constant in the definition of $S(\{\lambda\})$ matches with the introduced
pattern (\ref{4.1}).

We will need also the alternative representation of the 'measure' in
(\ref{4.1}) when the constrained sum
$\{\lambda\}\in{[{\bf Z}^{\oplus N}/S(N)}]$ over the
strictly decreasing integers is identically (for $q>0$) transformed 
into the unconstrained sum over the $independent$ integers
$\{\lambda\}\in{{\bf Z}^{\oplus N}}$
\be
\sum_{\{\lambda\}\in{[{\bf Z}^{\oplus N}/S(N)}]}~e^{-S(\{\lambda\})}=
\frac{1}{N!}
\sum_{\{\lambda\}\in{{\bf Z}^{\oplus N}}}~e^{-S(\{\lambda\})}~,
\label{4.1e}
\ee
where $S(\{\lambda_{i}\})=S(\{\lambda_{\sigma(i)}\}),~\sigma\in{S(N)}$.
It renders manifest the 'built in' Weyl group of the $S(N)$-permutations, so
that the constraint $\lambda_{i}>\lambda_{i+1}$ can be reinterpreted as the
'fixing' of the Weyl symmetry. As a result, the specific scaling (\ref{4.22})
can be viewed in fact as a consequence of the $S(N)$-invaraince of
$S(\{\lambda\})$ augmented by
the property that $e^{-S(\{\lambda\})}$ vanishes (owing to (\ref{4.3b})) on
$any$ boundary of the Weyl chamber, i.e. for
$\forall{\{\lambda_{i}=\lambda_{j},~i\neq{j}\}}$.

\section{The Continuum Limit and Large $N$ PTs.}
\setcounter{equation}{0}

The general idea of the continuum limit in lattice theories is
captured by the well-known intuitively transparent condition. Namely, the
properly introduced correlation length must tend to infinity (in the units of
the lattice spacing) so that the discrete space-time is 'smoothed-out', for
the low-energy theory, into the continuum manifold. In the lattice gauge
systems, the relevant coupling constant(s) should be adjusted in such a way
that the Wilson loop averages $<W_{C}(U)>$ undergo infinitely small changes
when the contour $C$ is deformed $microscopically$ (i.e. at the scales of the
$lattice$ cut off).

Although in a general case this criterion is rather difficult to
implement analytically, the specific structure of the lattice gauge theories
suggests a simpler alternative. It is intuitively clear that the infinite
correlation length is supposed to entail the $localization$ (modulo
(\ref{2.9}) and the gauge transformations) of the link-variables
$U_{\rho}({\bf z})$ in the infinitesimal vicinity of the
group-unity element $\hat{1}$. To be more specific, let us fix first the
'maximal tree' gauge \cite{Dr&Zub} putting $U_{\rho}({\bf z})=\hat{1}$ on
a largest possible $tree$ (made of the links) which by definition does not
contain nontrivial 1-cycles.
Then, introducing the quantum fluctuations $A^{ab}_{\rho}({\bf z})
=-iln[U^{ab}_{\rho}({\bf z})]$, the required localization can be formulated
in the large $N$ limit in the form
\be
\lim_{\tilde{g}^{2}N\rightarrow{0}}
<[A^{ab}_{\rho}({\bf z})]^{2}>\sim
{O(\tilde{g}^{2})}~\bmod (\ref{2.9})~,~\forall{a,b=1,...,N},
\label{7.2x}
\ee
where $\tilde{g}^{2}N\equiv
{\tilde{g}^{2}(\{g_{k}\})}N\rightarrow{0}$ is some $O(N^{0})$
functional (see below) of the relevant coupling constants $\{g_{k}\}$ that is supposed to
approach zero. The gauge-invariant representation of (\ref{7.2x}) evidently
reads as
\be
\lim_{N\rightarrow{\infty}}
\lim_{\tilde{g}^{2}N\rightarrow{0}} 
|\frac{1}{N}<tr[U(pl)]>-1|\sim{O(\tilde{g}^{2}N)}~,
\label{7.2xx}
\ee
where $U(pl)={U}_{\mu}({\bf x}){U}_{\nu}({\bf x+\mu})
{U}^{+}_{\mu}({\bf x+\nu}){U}^{+}_{\nu}({\bf x})$ stands for the holonomy
around an elementary plaquette in an arbitrary $\mu\nu$-plane.

Next, the scaling (\ref{7.2x}) can be translated into the following
(large $N$) constraint on the effective action $S(\{{A_{\rho}({\bf z})}\})$
for the 'gauge fields' $A^{ab}_{\rho}({\bf z})$. Recall that the latter can be
composed of not only all possible gauge invariant oprators but also
necessarily contains gauge noninvariant counter-terms (responsible for
the restoration of the Ward identities
following from the manifest gauge symmetry of the lattice gauge theory).
Employing the symbolic form (with $r(\{p\})\in{Y_{n}}$ standing for the
$S(n)$ irrep similarly to (\ref{4.1}))
\be
S(A)=\sum_{\{m\}}\sum_{n=2}^{+\infty} \sum_{r\in{Y_{n}}}
\beta(r(\{p\}),\{m\})~\prod_{k=1}^{n}
[tr((\partial)^{m_{k}}{A}^{k})]^{p_{k}},
\label{7.2m}
\ee
eq. (\ref{7.2x}) translates into the proper scaling of the defining
coefficients 
\be
\beta(r,\{m\})
\sim{O([N^{\frac{1}{2}}\tilde{g}]^{-n+\gamma(r,\{m\})}
N^{2-\sum_{k=1}^{n} p_{k}})}~~,~~\gamma(r,\{m\})\geq{0},
\label{7.2n}
\ee
where $A_{\rho}=0,~\forall{\rho},$ is supposed to be the global minimum
(modulo (\ref{2.9}) and gauge transformations) of the action (\ref{7.2m}).
In the case when the quadratic gauge invariant combination is $not$
unnaturally suppressed (i.e. $\gamma([2^{1}],2)=0$ but
$\beta([2^{1}],2)\neq{0}$), the formal (large $N$) 'continuum' limit
$A\rightarrow{0}$ of the lagrangian $L(A)$ yields
\be
\lim_{A\rightarrow{0}} \lim_{\tilde{g}^{2}N\rightarrow{0}}
L(A)\rightarrow{\frac{c}{{\tilde{g}}^{2}}
tr(F^{2}_{\mu\nu})}~~~;~~~c>0~,
\label{1.3b}
\ee
where $c\sim{O([\tilde{g}^{2}]^{0})}$ is some constant.
Remark that
the localization (\ref{7.2x}) does $not$ necessarily implies that in the large
$N$ WC limit $\tilde{g}^{2}N\rightarrow{0}$ the effective action
(\ref{7.2m}) becomes quadratic in $F^{2}_{\mu\nu}$. The $A\rightarrow{0}$
limit in eq. (\ref{1.3b}) may be omitted only when $\gamma(r,\{m\})>{0}$ for
$\forall{r}\in{Y_{n}},~n\geq{3}$.

Finally, in $D=4$ the
Renorm-Group analysis ensures the existence and $uniqueness$ of the
low-energy $YM$ theory renormalizable in the Dyson's sense (which is reassured
by asymptotic freedom). That is why one may expect that 
$D=4$ lattice gauge theories (\ref{7.2m}) belong to the
same universality class with the standard low-energy action
$\int d^{4}x ~tr[F^{2}_{\mu\nu}({\bf x})]/4g^{2}_{eff}$.

\subsection{The $\{U_{\rho}\rightarrow{\hat{1}}\}$ localization.}

In the particular case of gauge theories induced via (\ref{2.0z}) from the
eigenvalue-systems (\ref{6.25c}), the required localization of
$U_{\rho}({\bf z})$ is predetermined by the localization (\ref{6.70}) of the
link-variables $U_{\rho}$ in the reduced model (\ref{1.35}).
For definiteness, the effective 1-matrix system in the representation
(\ref{1.2b}) of (\ref{1.35}) is supposed to be restricted to (\ref{4.1}).
Let us demonstrate that the $\{U_{\rho}\rightarrow{\hat{1}}\}$ localization
(\ref{6.70}) holds true provided the saddle-point
values $\lambda^{(0)}_{i}=N\bar{\lambda}^{(0)}_{i}$ in (\ref{4.1}) approach
'infinity' according to the $scaling$-condition
\be
\lim_{N\rightarrow{\infty}}
\lim_{\tilde{g}^{2}N\rightarrow{0}} |\lambda^{(0)}_{i}|\sim
{O(N/[\tilde{g}N^{\frac{1}{2}}])}~~\Longleftrightarrow{~~
|\bar{\lambda}^{(0)}_{i}|\sim{O([\tilde{g}N^{\frac{1}{2}}]^{-1})}},
\label{7.1}
\ee
with a functional $\tilde{g}(\{g_{k}\})$ approaching zero.
To match with the localization (\ref{7.2x}), $\tilde{g}(\{g_{k}\})$ in the
above equation should be the same as in (\ref{7.2n}).

To begin with, we sketch under what conditions on the (rescaled according to
(\ref{4.1b}) coupling constants $b_{r(\{p\})}$ of the) effective model
(\ref{4.1}) the constraint (\ref{7.1}) is dynamically fulfilled. As for the
overall $O(N)$-scaling of $\{\lambda^{(0)}\}$, it has been demonstrated in
Section 3. This scaling is further augmented in the WC domain when $all$ the
parameters tend to zero: $\{b_{r}\rightarrow{0}\}$. In this case the large $N$
limit of the discrete $U(N)$ model (\ref{4.1e}) coincides with the
corresponding $hermitean$ model in the {\it eigenvalue}-representation
associated via substitution (\ref{4.1b})
\be
\lim_{N\rightarrow{\infty}}
\lim_{\{b_{r}\rightarrow{0}\}}
\sum_{\{\lambda\}} e^{-S(\{\lambda\}|\{g_{r}\})}=
\int_{-\infty}^{+\infty} \prod_{j=1}^{N} \frac{d\bar{\lambda}_{j}}{N^{-1}}
\prod_{i<k} (\bar{\lambda}_{i}-\bar{\lambda}_{k})^{2}
e^{-\tilde{S}(\{\bar{\lambda}\}|\{b_{r}\})},
\label{4.1d}
\ee
where $\tilde{S}(\{\bar{\lambda}\}|\{b_{r}\})=
S(\{N\bar{\lambda}\}|\{b_{r}N^{\gamma_{r}}\})+
2\sum_{i<k} ln|\bar{\lambda}_{i}-\bar{\lambda}_{k}|$.
One observes that the sums $\{\lambda\}\in{[{\bf Z_{N}}]^{\oplus N}}$ over
the $independent$ integer-valued $\{\lambda\}$-fields are
transformed into the corresponding $integrals$ over the real-valued variables
$\bar{\lambda}_{j}\sim{O(N^{0})}$. Complementary, in the WC domain the
constraints (\ref{3.46c}) in eq. (\ref{3.49}) are irrelevant and can be
omitted.

Defering the demonstration of these facts till
the next subsection, we employ the $WC$ representation (\ref{4.1d}) to
clarify the mechanism of the residual
$[\tilde{g}^{2}N]^{-1/2}$ scaling of $|\bar{\lambda}^{(0)}_{i}|$ and
relate the latter to the localization (\ref{6.70}). In the simplest case
of the $U(N)$ action (\ref{4.1}) with $M_{0}=1$, one simply
sets
\be
\sum_{R(\{\lambda\})} |dimR(\{\lambda\})|^{q}~
exp[{-g_{\tilde{r}_{0}}\sum_{i=1}^{N}(\lambda_{i}-\frac{N-1}{2})^{2}}]~~,~~
\tilde{g}^{2}/2=g_{\tilde{r}_{0}}~,
\label{4.1c}
\ee
where $g_{\tilde{r}_{0}}\sim{O(1/N)}$ and $\tilde{r}_{0}=[2^{1}]$. Rescaling
the eigenvalues $\bar{\lambda}_{j}=
[\tilde{g}N^{\frac{1}{2}}]^{-1}h_{j}$ and separating in
(\ref{4.1c}) the overall factor
$[\tilde{g}^{2}N]^{-\frac{qN(N-1)}{4}}$, we are left with
the $\tilde{g}^{2}N$-$independent$ theory of eigenvalues $h_{j}$
that altogether ensures (\ref{7.1}). In a general situation (\ref{4.1b}),
similar arguments hold true provided that $all$ $b_{r}$ scale to zero
as in (\ref{4.1d}). Choosing $b_{2k}=\liminf{[b_{r}]}$ as the smallest
$b_{r(\{p\})}$ in the subset of $r\in{Y_{2k}}$ for each particular
$k$, one is to equate $\tilde{g}N^{\frac{1}{2}}$ with the
$\liminf{[(b_{2k})^{\frac{1}{2k}}]}$ selected among all $k\leq{M_{0}}$. 

Now we are ready to make contact between the scaling (\ref{7.1}) and the  
localization $\{U_{\rho}\rightarrow{\hat{1}}\}$ of eq. (\ref{6.70}). Recall
that, according to eq.
(\ref{3.43}), the irreps $R$ in eq. (\ref{3.49}) represent the
irreps $\{R(\rho)\}$ entering the GLR fusion-rules (\ref{6.3p}). The latter
ensure that the scaling-condition (\ref{7.1}) is valid also for 
$\{\lambda^{(0)}(\mu\nu)\}$ parametrizing the SP irreps $\{R^{(0)}(\mu\nu)\}$
on which the large $N$ sum (\ref{1.35}), defining the reduced model, is
localized (after integration over $\{U_{\rho}\}$). Combining it with
(\ref{4.1d}), one concludes
that the eigenvalues $\omega_{j}(\rho)$ of
$U_{\rho}=\Omega_{\rho}~diag[e^{i\omega(\rho)}]~
\Omega_{\rho}^{+}$ are localized (modulo (\ref{2.9x})) in the domain where
$[S_{r}(\{U_{\rho}\})-S_{r}(\{\hat{1}\})]
\sim{O(N^{2}[\tilde{g}^{2}N]^{0})}$ when
$\tilde{g}^{2}N\rightarrow{0}$.
According to the Weyl character formula (\ref{4.3}), for any particular $j$ it
results in the large $N$ scaling complementary to (\ref{7.1})
\be
\lim_{N\rightarrow{\infty}}
\lim_{\tilde{g}^{2}N\rightarrow{0}} <\omega^{2}_{j}(\rho)>\sim{
O(\tilde{g}^{2}N)}~~\bmod [{\bf Z_{N}}]^{D}~,
\label{7.2}
\ee
where $[{\bf Z_{N}}]^{D}$ stands for (\ref{2.9x}). Alternatively, in terms of
the quantum fluctuations $A^{ab}_{\rho}=-iln[U^{ab}_{\rho}]$  the
condition
(\ref{7.1}) can be rewritten as
\be
\lim_{N\rightarrow{\infty}}
\lim_{\tilde{g}^{2}N\rightarrow{0}}
<[A^{ab}_{\rho}]^{2}>\sim{
O(\tilde{g}^{2})}~~\bmod [{\bf Z_{N}}]^{D}
\label{7.2b}
\ee
for any given $a,b=1,...,N$. Altogether, it justifies the required dynamical
localization (\ref{6.70}) in the reduced model (\ref{1.2b})/(\ref{4.1})
under the condition $\tilde{g}^{2}N\rightarrow{0}$.

Let us now return to the localization (\ref{7.2xx}) in the gauge theories
induced from the eigenvalue-systems (\ref{6.25c}) associated to the reduced
models satisfying (\ref{7.2b}). The pattern of the mapping
(\ref{2.0z}) (together with the saddle-point analysis of Section 2) suggests
that the functionals $\tilde{g}(\{g_{k}\})$ in eqs. (\ref{7.2x}) and
(\ref{7.2b}) are to be identified. To substantiate this identification, we
consider the large $N$ WC asymptotics
$\tilde{g}^{2}N\rightarrow{0}$ of the properly
normalized partitition function (PF)
\be
X^{(in)}_{L^{D}}=\int \prod_{\{\rho,{\bf z}\}}dU_{\rho}({\bf z})
exp[{-S(\{U_{\rho}({\bf z})\})-S(\{\hat{1}\})}]
\label{7.3m}
\ee
for a (induced) gauge theory on a cubic lattice with $L^{D}$ sites.
As the construction (\ref{7.3m}) excludes the factor
$e^{-S(\{\hat{1}\})}$ irrelevant for the contribution of the WC
perturbative series, in the limit ${\tilde{g}^{2}N\rightarrow{0}}$
the PF $X^{(in)}_{L^{D}}$ yields the leading order of the large $N$
$(\tilde{g}^{2}N)$-expansion.

Consider a generic gauge theory (\ref{7.2m}) with the localization
(\ref{7.2xx}) imposed through the pattern (\ref{7.2n}). Then the WC
asymptotics of (\ref{7.3m}) is predetermined by the
$\tilde{g}^{2}N$-scaling law of the measure i.e. active Haar link-integrations
remaining after a gauge fixing. This is simply because the
exclusion of the $e^{-S(\{\hat{1}\})}$-factor guarantees that the overall
action $[S(\{U_{\rho}({\bf z})\})-S(\{\hat{1}\})]$ is 
$\sim{O(N^{2}{[\tilde{g}^{2}N]}^{0})}$ when
$\tilde{g}^{2}N\rightarrow{0}$. As for the Haar measure scaling,
consider first a cubic $L^{D}$ lattice
with the free boundary conditions for $U_{\rho}({\bf z})$. One observes that
in the limit $L\rightarrow{\infty}$ the maximal tree gauge in the average
excludes at each site one link-integration out of the total amount $D$.
The latter is clear if one imposes ('almost' complete when
$L\rightarrow{\infty}$) temporal axial gauge keeping
$U_{D}({\bf z})=\hat{1},~\forall{{\bf z}}$. Each remaining active Haar
integration
merges, in the vicinity of $\hat{1}$, with the 'flat' Lebesgue measure
$\prod_{\rho=1}^{D-1}d^{N^{2}}A_{\rho}({\bf z})$. Summarizing, for
$L\rightarrow{\infty}$ the
localization (\ref{7.2x}) results in the following large $N$ WC asymptotics
\be
\lim_{N\rightarrow{\infty}}
\lim_{\tilde{g}^{2}N\rightarrow{0}}
X^{(in)}_{L^{D}}(\tilde{g})=[C~\tilde{g}N^{\frac{1}{2}}]
^{(D-1) N^{2}L^{D}}~~,~~C>0~,
\label{7.4b}
\ee
where the $\int dU=1$ normalization of the Haar measure
is used, and $C$ is a model dependent constant. Finally, one can demonstrate
that the above pattern remains valid for a finite $L$ provided the
choice of the periodic boundary conditions for $U_{\rho}({\bf z})$.
In the specific case of the Wilsonian lattice action (where in eq.
(\ref{7.2n}) $\gamma(n,m)=n-2$ so that (\ref{7.4b}) results from a gaussian
integration) the above asymptotics was derived in \cite{Creutz}.

Summarizing, to justify the identification of the functionals
$\tilde{g}(\{g_{k}\})$ in eqs. (\ref{7.2x}) and (\ref{7.2b}), one is
to prove that eq. (\ref{7.4}) holds true for the gauge theories induced
from the reduced models constrained by (\ref{7.1}). To begin with, according
to the mapping (\ref{2.0z}), the factor $e^{-S(\{\hat{1}\})}$ can be rewritten
as the partitition function (PF) $\tilde{X}^{(a)}_{L^{D}}$ of the auxiliary
model. The latter is to be computed in terms of the plaquatte-factor
(\ref{6.25c}) where the substitution
\be
U_{\rho}({\bf z})\rightarrow
{\tilde{G}^{+}({\bf z})\tilde{G}({\bf z+\rho})}
\label{6.25v}
\ee
is performed. Consequently, the properly normalized PF of the induced gauge
theory can be representaed as the ratio
\be
X^{(in)}_{L^{D}}=X_{L^{D}}/X^{(a)}_{L^{D}}
\label{7.4v}
\ee
where $X_{L^{D}}$ and $X^{(a)}_{L^{D}}$ are the PFs (both normalized akin to
(\ref{7.3m})) associated to the eigenvalue-system (\ref{6.25c}) and the
auxiliary model defined through (\ref{6.25v}) respectively.

As for the large $N$ limit of $X_{L^{D}}$, the correspondence (\ref{2.10})
allows to express it as $L^{D}th$ power of the PF $X_{r}$ of the reduced 
model (\ref{1.35}). In turn, the localization (\ref{7.2b}) results in
the $power$-like asymptotics
\be
\lim_{N\rightarrow{\infty}}
\lim_{\tilde{g}^{2}N\rightarrow{0}}
X_{r}(\{g_{r}\})=
[{B} ~\tilde{g}N^{\frac{1}{2}}]^{D N^{2}}~~,~~{B}>0~,
\label{7.4}
\ee
which is deduced similarly to (\ref{7.4b}) (while $\int dU=1$ is
presumed).
Complementary, one easily observes that the localization (\ref{6.70})
in eigenvalue-system (\ref{6.25c}) predetermines that in the auxiliary model
(\ref{6.25v}) the $SU(N)$ field $\tilde{G}({\bf z})$ is localized (modulo 
(\ref{2.9})) in the
vicinity of $\hat{1}$ as well. Consequently, akin to (\ref{7.4}) one obtains
\be
\lim_{N\rightarrow{\infty}}
\lim_{\tilde{g}^{2}N\rightarrow{0}}
X^{(a)}_{L^{D}}(\{g_{r}\})=
[\tilde{B}~\tilde{g}N^{\frac{1}{2}}]^{L^{D} N^{2}}~~,~~\tilde{B}>0~.
\label{7.4vv}
\ee
Combining all the pieces together, we finally arrive at the required
asymptotics (\ref{7.4b}) with $\tilde{g}(\{g_{k}\})$ being identical to the
functional which enters the scaling (\ref{7.1}) in the associated
reduced model.

\subsection{The auxiliary 'continuum' limits and $N\rightarrow{\infty}$ PTs.}

The 1-matrix representation (\ref{3.49}),(\ref{4.1}) of the
large $N$ PF (\ref{6.3p}) allows, for the first time, to address analytically
the $D>2$ pattern of the phase transitions (PTs) in the $SU(N)$ gauge theories
induced from the eigenvalue-systems (\ref{6.25c}).
In particular, the asymptotics (\ref{7.4b}) of $X^{(in)}_{L^{D}}$ (and
similar power-like asypmtotics of $\tilde{X}^{(in)}_{L^{D}}\rightarrow{
(\tilde{X}_{r})^{L^{D}}}$) is characteristic of the phase
naturally refered to as the weak-coupling (WC) one. The latter is associated
to a connected vicinity of the WC domain $\{b_{r}\rightarrow{0}\}$ of the
parameters defining the effective 1-matrix system (\ref{4.1}). On the other
hand, the strong-coupling (SC) phase (associated to a connected vicinity of
$\{b_{r}\rightarrow{\infty}\}$) can be shown to correspond to a completely
different pattern of $\tilde{X}^{(in)}_{L^{D}}(\{g_{k}\})$. To take the
simplest
example, consider $D=2$ (where the constraints (\ref{3.46c}) are trivial) and
choose the simplest 1-matrix model (\ref{4.1c}) with $q=2$. As it is known
\cite{Gr&Tayl}, the large $N$ free energy of the latter model can be expanded in
the SC series
\be
-\lim_{N\rightarrow{\infty}} ln[\tilde{X}_{1}(\tilde{b})]=
N^{2}\sum_{n=1}^{+\infty} \sum_{m=0}^{M(n)} f(n,m)~\tilde{b}^{2m}
exp[{-{n\tilde{b}^{2}}/{2}}]
\label{7.4xx}
\ee
which reproduces the correct answer in a connected vicinity of $\tilde{b}=
\tilde{g}^{2}N\rightarrow{\infty}$.

The reason for the mismatch between the WC and SC large $N$ patterns
resides in
the phase transition(s) which are ubiquitous \cite{Dougl&Kaz,Gr&Witt} in the
lattice gauge systems. As we will see, in the case of the induced gauge
theories, it implies that the large $N$ continuum limit (CL) encoded by
(\ref{7.2m})-(\ref{1.3b}) is
accompanied by a few $auxiliary$ 'continuum' limits in the effective 1-matrix
system (\ref{3.49}). The latter are characterized by the
transformation of a few relevant discrete spaces into the corresponding
continuum manifolds.

As previously, we concentrate on the simplest case when the 1-matrix model in
(\ref{3.49}) reduces to the selected family (\ref{4.1}). The appropriate
framework for our analysis is provided by the method based on
the saddle-point equations (originally introduced in \cite{Brez}) in a
particular realization close to \cite{Rus,Dougl&Kaz}.
First of all, let us demonstrate that in the
large $N$ limit the $SU(N)$ $D>2$ constraints (\ref{3.46c}) in eq.
(\ref{3.49}) are irrelevant and can be omitted. The easiest way to see it is
to notice first that for a selfdual $SU(N)$ irrep
\be
n(R)|_{R=\bar{R}}=N(\lambda_{1}-N+1)/2
\label{3.46x}
\ee
so that, choosing $N\in{2(D-1){\bf Z_{>0}}}$, we satisfy
(\ref{3.46c}) for $\forall{\{\lambda\}}$. Presuming the 'smoothness' of the
large $N$ limit, one arrives at the required conclusion. Actually, employing
the 'regularized' Poisson resummation formula
(\ref{3.49b}), the irrelevance of (\ref{3.46c}) in the $SU(N)$ models can be
explicitly proven  with the help of the formalism we now focus on to handle
the effective 1-matrix system (\ref{4.1}).

As we demonstrated in Section 3.2, the adjustment (\ref{4.1b}) of the
$\{g_{k}\}$-scaling is sufficient to ensure that the finite $N$ sum
(\ref{4.1}) is accumulated by the configurations $\{\lambda_{j}\}$ possessing
a 'smooth' large $N$ limit $both$ in the index $j$-space $and$ in the
'base' $\{\lambda\}\in{{\bf Z}^{\oplus N}}$-space. Then, in the computation of
the leading $O(N^{2})$ order of $-ln(\tilde{X}_{1})$ the large $N$ sum
(\ref{4.1})
can be substituted by the corresponding 'path-integral'
\be
\tilde{X}_{1}=\int D\bar{\lambda}(t)~
exp({-N^{2}{S}_{eff}[\bar{\lambda}(t)]}~)
\label{4.18}
\ee
\be
\bar{\lambda}(1-\frac{i}{N})\equiv{\frac{\lambda_{i}}{N}=\bar{\lambda}_{i}}
~~~,~~~N^{2}{S}_{eff}[\lambda(t)]\equiv{S(\{\lambda\})},
\label{4.18b}
\ee
over the '$continuum$' variables $\bar{\lambda}(t)$ evidently obeying the
inequality \cite{Dougl&Kaz} following from (\ref{2.3})
\be
\lambda_{i}-\lambda_{i+1}\rightarrow
\frac{d\bar{\lambda}(t)}{dt}
\geq{1}~~,~~t=(1-i/N)\in{[0,1]}~.
\label{4.19}
\ee
If the parameters $\{g_{k}\}$ are such that the solution
$\bar{\lambda}^{(0)}(t)$ of the saddle-point equations
\be
\frac{\delta S_{eff}[\bar{\lambda}(t)]}
{\delta \bar{\lambda}(t)}=0
~~,~~\bar{\lambda}^{SU(N)}(0)=0~,
\label{4.21}
\ee
fulfils $d\bar{\lambda}^{(0)}(t)/dt\geq{1}$ for $\forall{t}\in{[0,1]}$, the
$U(N)$ pattern of (\ref{4.18}) reduces to that of the large $N$ $hermitean$
model associated to (\ref{4.1}) via the correspondence (\ref{4.1d}). As we
will prove below, this reduction occurs in the $WC$ phase when the set
$\{g_{r}\} \cong{\{b_{r}\}}$ belongs to a connected vicinity of
$\{b_{r}\rightarrow{0}\}$. In the corresponding hermitean
model, the $1/N$ quasiclassical expansion is organized in the background of
the saddle-point solution(s) $\bar{\lambda}^{(0)}(t)$. Since
$S(\{\lambda\})\sim{O(N^{2})}$, the quantum fluctuations of the continuum
$\lambda(t)$-variables can be neglected in the large $N$ limit 
\be
\lim_{N\rightarrow{\infty}}\tilde{X}_{1}=
exp(-N^{2}{S}_{eff}[\bar{\lambda}^{(0)}(t)])
\label{4.23}
\ee
and the $leading$ order of the semiclassical and that of the $1/N$ expansions
$coincide$. Remark that, in the case associated to the $SU(N)$ system,
we are to impose the additional 'boundary' condition (\ref{4.21}). The latter
matches with the translational invariance $\lambda_{i}\rightarrow
{\lambda_{i}+m}$ which must be present in the (${\bf Z_{2}}$-invariant)
$SU(N)$ action $S(\{\lambda^{SU(N)}\})$ defined by eq. (\ref{4.1}).

One can show that the $SU(N)$ solution of (\ref{4.21}),(\ref{4.1}) does
exists (provided the subset $\{g_{r},~
r\in{Y_{2M_{0}}}\}$ has the signs consistent with the convergence of the
$\{\lambda\}$-series), being unique and ${\bf Z_{2}}$-invariant. In the
$U(N)$-case the above $m$-translations (of $\lambda_{i}$) generate the
'quasizero' mode (to be factorized out in the analysis of the $O(N^{2})$-order
of $-ln[\tilde{X}_{1}]$). In effect, it suffices to retain the (unique)
${\bf Z_{2}}$-invariant $U(N)$ solution which satisfies 
\be
h^{(0)}(t)=-h^{(0)}(1-t)~~,~~h(t)=\bar{\lambda}(t)-1/2~,
\label{4.55v}
\ee
in terms of the shifted function $h(t)$. The explicit construction is easier
to perform reformulating (\ref{4.1}) in terms of the spectral density
associated to $h^{(0)}(t)$
\be
0\leq{\rho(\eta)}=\int_{0}^{1} dt \delta(\eta-h^{(0)}(t))=
(dt/dh^{(0)}(t))|_{h^{(0)}(t)=\eta}\leq{1}~,
\label{4.55}
\ee
constrained by $\int d\eta \rho(\eta)=1$. Remark that, in the case
corresponding to the $SU(N)$ model (\ref{4.1}), one is
to impose additionally that $\rho^{SU(N)}(\eta)=0$ if $\forall{\eta}<-1/2$
to match with the condition (\ref{4.21}). Note also that condition
(\ref{4.19}) results in the constraint $\rho(\eta)\in{[0,1]}$.

To take the simplest example, consider the $U(N)$ subvariety of (\ref{4.1})
when the factor in the exponent $e^{-V(R)}$ can be rewritten in terms of
$h(t)$ in the 'local' form
\be
\frac{V(R(\{\lambda\}))}{N^{2}}=\int dt~V(h(t))~~~,~~~
V(\eta)=\sum_{m=1}^{M_{0}} \frac{[~\tilde{b}_{2m}~\eta~]^{2m}}{2m}~,
\label{4.55vv}
\ee
corresponding to $p_{2k}=1,~p_{2k-1}=1,~{1\leq{k}\leq{M_{0}}}$. In this
case, the SP equation for $\rho(\eta)$ reads
\be
\frac{1}{q}\frac{dV(\eta)}{d\eta}=P\int d\phi~\frac{\rho(\phi)}{\eta-\phi}~~~~
,~~~~\rho(\phi)=\rho(-\phi),
\label{4.55w}
\ee
where $P$ stands for the principle value of the integral over $\phi$, and
the additional reflection-invariance condition selects the solution
associated to a selfdual SP irrep $R^{(0)}=\bar{R}^{(0)}$.
There exists a standard algorithm to solve (\ref{4.55w}) for an arbitrary
potential (\ref{4.55vv}). It can be found e.g. in the third reference of
\cite{Rus} and will not be reviewed here. We note only that, in the case of
the simplest $M_{0}=1$ $U(N)$ model (\ref{4.1c}), the solution $\rho(\eta)$ of
(\ref{4.55w}) assumes the pattern of the Wigner semicircle law
\be
\rho(\eta)=\frac{\tilde{b}^{2}}{q\pi}
\sqrt{\frac{2q}{\tilde{b}^{2}}-\eta^{2}}~~,~~\tilde{b}\equiv{\tilde{b}_{2}}~,
\label{4.55x}
\ee
which allows to reproduce the original descrete sum (\ref{4.1}) for
$(2\tilde{b}^{2}/q)<\pi^{2}$ (when $\rho(\eta)<1$ in the whole admissible
domain of $\eta^{2}\leq{{2q}/{\tilde{b}^{2}}}$).

Next, in the domain of $\{g_{r}\}$ where $d\bar{\lambda}^{(0)}(t)/dt<{1}$
in a set of 'windows' $t\in{\cup_{k}~[\tilde{t}_{a(k)},\tilde{t}_{b(k)}]}$,
the above strategy has to be refined. To be more specific, consider again the
simplest $M_{0}=1$ $U(N)$ model (\ref{4.1c}) (analysed at length in
\cite{Dougl&Kaz}). When
$\tilde{b}^{2}>{\tilde{b}_{cr}^{2}}=q\pi^{2}/2$, the solution
$\bar{\lambda}^{(0)}(t)$ is
located partially outside the admissible domain (\ref{4.19}) as it is clear
from (\ref{4.55x}). To reproduce the original discrete model (\ref{4.1c}) with
$\tilde{b}^{2}>\tilde{b}_{cr}^{2}$, in (\ref{4.23}) one is to
substitute $\bar{\lambda}^{(0)}(t)$ by the 'closest' $boundary$ configuration
$\bar{\zeta}^{(0)}(t)$. The latter is to be found presuming that
\be
\frac{\bar{\zeta}^{(0)}(t)}{dt}=1~~~,~~~t\in{[t_{1},t_{2}]}~,
\label{4.21b}
\ee
where the $[\bf Z_{2}]$-sefduality prescribes that $-t_{1}+1/2=t_{2}-1/2$.
The choice (\ref{4.21b}) implies simply that, for
the indices $j$ corresponding to the window
$[t_{1},t_{2}]$, the number $n_{j}$ of boxes in the
associated $jth$ rows of $Y_{n}^{(N)}$ (constrained by $n_{i}\geq{n_{i+1}}$)
assumes the '$boundary$' values $n_{i}=n_{i+1}=..=n_{N-i}=0$. After
substitution of
(\ref{4.21b}) into the effective action (\ref{4.18b}), one obtains a
(modified compared with (\ref{4.21})) saddle-point equations for
$\bar{\zeta}^{(0)}(t)$ in the remaining domain $t\in{[0,t_{1}]
\cup [t_{2},1]}$.

Practical computations are easier to perform translating
the above construction in terms of the spectral density (\ref{4.55}) to be
deformed $\rho(\eta)\rightarrow{1}$ on the interval
$\eta\in{[\eta_{1},\eta_{2}]}$ corresponding to (\ref{4.21b}).
The 'turning on'
(\ref{4.21b}) of the constraint (\ref{4.19}) (being a nonanalytical
procedure) results in the third order phase transition 
\cite{Dougl&Kaz}. Note also that in the $M_{0}>1$ case of (\ref{4.1})
there may be multicritical pattern with a few phase transitions associated to
opening or closing of a new 'window' $[t_{a(l)},t_{b(l)}]$ (see e.g. the
third ref. in \cite{Rus}).

Next, the above picture of PTs allows to reinterprete the mismatch between
the large $N$ WC and SC patterns of the partitition function
$\tilde{X}_{L^{D}}$ (eqs. (\ref{7.4b}) and (\ref{7.4xx}) respectively)
from the viewpoint of the auxiliary continuum limits. To begin with,
according to the identification 
(\ref{4.19}) the $N\rightarrow{\infty}$ limit of the discrete index-space
$\{i=1,...,N\}$ is the continuum 'time'-manifold $t$. Outside the WC domain,
the large $N$ system (\ref{4.18}) still retains the remnant
of the discreteness (associated to the base
${\bf Z}^{\oplus N}$-space) which is encoded in the constraint (\ref{4.19}).
The subtlety is that the latter constraint is relevant
despite the fact that each particular large $N$
sum over a given $j$-species $\lambda_{j}\in{\bf Z}$ merges with the integral
over the real-valued variable. Finally, let us show that in the WC limit
$\{b_{r}\rightarrow{0}\}$ the solution $\bar{\lambda}^{(0)}(t)$ doesn't
violate condition (\ref{4.19}). In other words, all the involved discrete
spaces
merge with the associated continuum manifolds which, in particular, implies
the validity of the large $N$ reduction (\ref{4.1d}) of (\ref{4.1}) in the
whole $N\rightarrow{\infty}$ WC phase. Indeed, the substitution (\ref{4.1d})
can be justified by going over to the effective theory of the
$\chi_{j}=[\tilde{g}N^{\frac{1}{2}}]\bar{\lambda}_{j}$ fields. Having adjusted
the proper scaling (\ref{4.1b}),(\ref{7.1}) (i.e.
$\tilde{g}^{2}N$-independence of the large $N$ $\chi_{j}$-system in the
WC limit), one
obtains
\be
d\bar{\lambda}(t)/dt=[\tilde{g}N^{\frac{1}{2}}]^{-1}
d\chi(t)/dt\sim{O(1/[\tilde{g}N^{\frac{1}{2}}])}>>1
\label{4.21x}
\ee
which substantiates the WC equivalence of (\ref{4.1}) with the associated
hermitean model.

In conclusion, it takes a minor modification of the above analysis
to retain explicitly the $D>2$ $SU(N)$ constraints (\ref{3.46c}) built into
the effective 1-matrix model (\ref{1.2b}). Employing the representation
(\ref{3.49}) of the latter, one can demonstrate that the free energy
$N^{2}F_{eff}(p,\bar{p})$ (evaluated prior to the sum over
$p,\bar{p}\in{\bf Z}$) assumes the form
\be
\lim_{N\rightarrow{\infty}}
e^{-N^{2}[F_{eff}(p,\bar{p})-F_{eff}(0,0)]}=
\lim_{\varepsilon\rightarrow{0}}~ 
e^{-\varepsilon p^{2}}~
\delta_{\bar{p},0}~\sum_{k\in{\bf Z}} \delta_{p,[D-1]k}~.
\label{7.20}
\ee
As a result, the localization $\{\bar{p}=0;~p=0 \bmod [D-1]{\bf Z}\}$ proves
that the constraints (\ref{3.46c}) can be safely omitted in (\ref{1.2b}),
(\ref{3.49}) when $N\rightarrow{\infty}$.

Summarizing, the large $N$ continuum limit (CL) in the
proposed induced lattice gauge theories is accompanied by the two
auxiliary 'CLs'. The latter, being predetermined (in the reduced system
(\ref{3.49})) by the correspondence (\ref{4.22}) and the scaling
(\ref{7.1}), are associated to the $N\rightarrow{\infty}$
transformation of the discrete $j$-index space and $\{\lambda\}
\in{{\bf Z}^{\oplus N}}$ space into the corresponding continuum manifolds.
In turn, it foreshadows the irrelevance of the $D>2$ constraint (\ref{3.46c})
in the limit $N\rightarrow{\infty}$. Together with the large
$N$ reduction (\ref{2.10}), it results in the substantial simplification
of the large $N$ WC analysis.
As for the physical interpretation of the discussed phase pattern, it can
be understood relating the large $N$ PTs to the issue of the validity of the
WC and the SC
expansions in the associated WC $\{b_{r}\rightarrow{0}\}$ and the SC
$\{b_{r}\rightarrow{\infty}\}$ domains respectively. We expect also that,
among the phase transitions (passing from the WC to the SC limits), there
is one associated to the condensation of the $microscopic$ lattice monopoles
which are always present in the lattice gauge theories.

\section{Remarks about the D=2 case.}
\setcounter{equation}{0}

To begin with, consider the $D=2$ large $N$ average of some
nonself-intersecting Wilson loop (\ref{2.13}) for which a closed
expression can be derived. The appropriate method is a minor modification
of the combinatorial approach due to Migdal \cite{Migd75} (see also
\cite{Rus,Witt91}) developed in the context of the ordinary $2d$ gauge
theories.

Owing to the periodic boundary conditions the
$2d$ $L^{2}$-plane in question is made, topologically, into a discretized
2-tora. Let
$L\rightarrow{\infty}$ so that any finite-size loop $C$ does not coincide
with one of the two uncontractible cycles defining the tora. Consequently,
cutting along the contour $C$, this tora is decomposed into the two disjoint
connected componenets: a disk with area $A$ and the complementary disk with
one handle of the area $L^{2}-A$. To apply cutting-gluing technique,
first one is to introduce (as in \cite{Witt91}) the so-called
'disc-amplitudes' $Z_{G}(\tilde{A},L_{C}|\{U({\bf z_{k}})\})$. The latter is
defined
as the partitition function of the $D=2$ eigenvalue-theory (\ref{6.25c}) on a
disk of the area $\tilde{A}$ with a given number of handles $G$. The
boundary-contour $C$ is supposed to consist of $L_{C}$ $\rho_{k}$-links, each
being endowed with a $free$ boundary conditions introduced via the associated
link-variables $U_{\rho_{k}}({\bf z_{k}}),~k=1,...,L_{C}$.

A particular term of each defining factor
(\ref{6.25c}), being refered to a $\mu\nu$-plaquette based at ${\bf z}$,
assigns to this plaquette one of the $SU(N)$ irreps $R_{\mu\nu}({\bf z})
\equiv{R({\bf z})}$. The specifics in the computation of the $2d$
disc-amplitudes is that $Z_{G}({A},L_{C}|\{U({\bf z_{k}})\})$ can be evaluated 
with the help of the orthonormality condition for the Lie group characters
\be
L^{(2)}_{R_{1}|R_{2}}=\int dV~\chi_{R_{1}}(V)~\chi_{R_{2}}(V^{+})=
\delta_{R_{1},R_{2}},
\label{8.1}
\ee
corresponding to the simplest GLR coefficient of the second order while the
higher order $L^{(p)}_{R_{+}|\{R_{k}\}},~p\geq{3},$ are $not$ involved. Given
the amplitude associated to a particular connected base-surface, eq.
(\ref{8.1}) prescribes that $R_{\mu\nu}({\bf z})$ is ${\bf z}$-$independent$.
As a result, thus introduced disk-amplitude is readily computed in the
$G$-$independent$ form
\be
Z_{G}({A}|\{U({\bf z_{k}})\})=
\sum_{R} e^{-S(R){A}} \prod_{\{{\bf z_{k}}\in{C}\}}
\chi_{R}({U}_{\rho_{k}}({\bf z_{k}}))
\label{8.2}
\ee
where $\{{\bf z_{k}}\}$ stands for the set of $L_{C}$ sites on the boundary
$C$. As a cross-check, the $D=2$ partitition function $\tilde{X}_{L^{2}}$ can
be easily rewritten in terms of $Z_{G}({A}_{q}|\{U({\bf z_{k}})\})$
(where $A_{1}=A$, $A_{2}=L^{2}-{A}$)
\be
\tilde{X}_{L^{2}}=
\int \prod_{\{{\bf z_{k}}\}} dU_{\rho_{k}}({\bf z_{k}})~
Z_{0}(A_{1}|\{U^{+}({\bf z_{k}})\})~
Z_{1}(A_{2}|\{U({\bf z_{k}})\})
\label{8.4}
\ee
which, owing to (\ref{8.2}) and (\ref{3.45}), results in
\be
\lim_{N\rightarrow{\infty}} \tilde{X}_{L^{2}}=
\lim_{N\rightarrow{\infty}}\sum_{R} e^{-S(R)L^{2}}=
[~\sum_{R} e^{-\frac{S(R)}{2}}~]^{2L^{2}}
\label{8.5}
\ee
that matches with the $D=2$ case of (\ref{6.3p}) according to (\ref{8.1}).
Remark that the same expression comes out in the case of the $L^{2}$-surface
of an $arbitrary$ genus $G$.

Next, similarly to the ordinary $2d$ lattice gauge theories, the Wilson loop
average in the $2d$ system (\ref{6.25c}) reads
\be
<W^{R_{0}}_{C}>=
\int \prod_{\{{\bf z_{k}}\}} dU_{\rho_{k}}({\bf z_{k}})
Z_{0}(A_{1}|\{U^{+}({\bf z_{k}})\})
Z_{1}(A_{2}|\{U({\bf z_{k}})\})
\frac{\chi_{R_{0}}(U_{C})}{\tilde{X}_{L^{2}}},
\label{8.3}
\ee
where $U_{C}=\prod_{\{{\bf z_{k}}\}}U_{\rho_{k}}({\bf z_{k}})$, $R_{0}$ is the
representation of the holonomy.
Evaluating (\ref{8.3}) one is to
take advantage
of the fact that $Z_{G}({A}|\{U({\bf z_{k}})\})$ is independent of the
'angular' variables $\Omega({\bf z_{k}})$. Equivalently, it can be represented
as the invariance of $Z_{G}(\tilde{A}|\{U({\bf z_{k}})\})$ with respect to
the $[U(N)]^{\oplus L_{C}}$ conjugations
\be
U_{\rho_{k}}({\bf z_{k}})\rightarrow
{V_{\rho_{k}}^{+}({\bf z_{k}})U_{\rho_{k}}({\bf z_{k}})
V_{\rho_{k}}({\bf z_{k}})}~~~,~~~V_{\rho_{k}}({\bf z_{k}})\in{U(N)}~,
\label{8.6}
\ee
(where $k=1,...,L_{C}$) inherited from the symmetry (\ref{2.50}) of the
factor (\ref{6.25c}). Let us perform the substitution (\ref{8.6})
as a change of the variables in eq. (\ref{8.3}) that leaves the measure
invariant. Integrating over $V_{\rho_{k}}({\bf z_{k}})$ with the Haar measure
and employing the identity
\be
\int \prod_{\{{\bf z_{p}}\}} dV({\bf z_{p}})
\chi_{R_{0}}(\prod_{\{{\bf z_{k}}\}}
V^{+}({\bf z_{k}})U_{\rho_{k}}({\bf z_{k}})
V({\bf z_{k}}))=
\frac{\prod_{\{{\bf z_{k}}\}}\chi_{R_{0}}(U_{\rho_{k}}({\bf z_{k}}))}
{[dimR_{0}]^{L_{C}-1}},
\label{8.7}
\ee
one trades the factor
$\chi_{R_{0}}(\prod_{\{{\bf z_{k}}\}}U_{\rho_{k}}({\bf z_{k}}))$
for the properly normalized product of the characters (akin to (\ref{1.31})).
At the same time, the rest of the integrated in (\ref{8.3}) expression is left
intact. Remark that in eq. (\ref{8.7}) the $[dimR_{0}]^{1-L_{C}}$
normalization can be deduced
making the particular choice of the arguments: $U_{\rho_{k}}({\bf z_{k}})=
\hat{1},~\forall{k}$.

After $\{dV({\bf z_{p}})\}$ integrations, a generic 
loop average (\ref{2.13}) in the eigenvalue systems can be expressed 
in terms of the GRL coefficients (\ref{6.3a}). For a nonself-intersecting
contour $C$ one obtains
\be
<W^{R_{0}}_{C}>=\frac{1}{\tilde{X}_{L^{2}}}\sum_{R_{1},R_{2}}
e^{-S(R_{1})A-S(R_{2})(L^{2}-A)}~
\frac{[L_{R_{1}|R_{2},R_{0}}]^{L_{C}}}
{[dimR_{0}]^{L_{C}-1}}
\label{8.8}
\ee
which complements the GLR pattern of the partitition function. Observe that
the same GLR pattern (\ref{2.13b}) of $<W^{f}_{C}>$ immediately follows from
(\ref{8.7}) in $any$ $D\geq{2}$. Finally,
$<W^{R_{0}}_{C}>$ can be rewritten in a more conventional form of the average
\be
<W^{R_{0}}_{C}>=\frac{1}{\tilde{X}_{L^{2}}}\sum_{R}
e^{-L^{2}S(R)}~B_{L_{C}}(R|R_{0},A)\equiv{<B_{L_{C}}(R|R_{0},A)>_{R}}
\label{8.9}
\ee
performed with the weight $e^{-L^{2}S(R)}$, and we have introduced
\be
B_{L_{C}}(R|R_{0},A)=\sum_{R_{1}}e^{-[S(R_{1})-S(R)]A}~
\frac{[L_{R_{1}|R,R_{0}}]^{L_{C}}}
{[dimR_{0}]^{L_{C}-1}}~.
\label{8.10}
\ee
Remark that eqs. (\ref{8.9}),(\ref{8.10}) remain valid in the case of a
generic nonself-intersecting contour $C$ on a discretized $2d$ surface of
an $arbitrary$ genus (provided $C$ is not homotopic to any of defining
uncontractible cycles).

Building on the results of \cite{Boulat} (obtained in the context of the
continuum $YM_{2}$ on a $2d$ sphere), one can derive the integral
representation for (\ref{8.9}) in terms of the associated spectral
density (\ref{4.55}). For simplicity, we consider the option when $e^{-S(R)}$
belongs to the $U(N)$ subvariety (\ref{4.55vv}) of the family (\ref{4.1}).
Also, we restrict our attention to the segment of the $WC$ domain of  
$\{{g}_{r}\}\cong{\{{b}_{r}\}}$ where the spectral
density is less than $1/2$ for all admissible $\eta$. Then, as
it is demonstrated in Appendix B, the large $N$ average $<W^{f}_{C}>$ of a
nonself-intersecting loop $C$ in the fundamental representation assumes the
form
\be
<W^{f}_{C}>=N^{2-L_{C}}\int d\eta~ \rho(\eta)~
\left(\frac{sin(\pi\rho(\eta))}{\pi\rho(\eta)}\right)^{{qA}}~,
\label{A.14v}
\ee
where $\rho(\eta)<1/2$ is the ${\bf Z_{2}}$-invariant solution of the SP eq.
(\ref{4.55w}). We postpone the interpretation of the unconventional
$N^{2-L_{C}}$-scaling of the average (\ref{A.14v}) till the next section and
now turn to the issue of the continuum limit (CL).

\subsection{Identification of the universality class.}

To study the CL of (\ref{A.14v}) directly, one is to send $L^{2},A$ to
infinity keeping $A/L^{2}$ finite. Simultaneously, the coupling constant
$\tilde{b}^{2}={g}^{2}_{r}N$ (entering (\ref{7.1})) should approach the
critical value $\tilde{b}^{2}=0$
(corresponding to the localization (\ref{6.70})). It is to be performed in
accordance with such scaling-law which ensures a $finite$ physical string
tension $\sigma$ as $\tilde{b}^{2}\rightarrow{0}$. Let
the total area of the closed $2d$ surface in question, being measured in the
units of $\sigma^{-1}$, is adjusted to be finite as well. Take the $G=0$ case,
the $2d$ sphere, where the free energy of $continuum$ $YM_{2}$ is
$\sim{O(N^{2})}$. Then thus defined $\tilde{b}^{2}\rightarrow{0}$
limit of the (area-dependent part of) the $G=0$
amplitude (\ref{A.14v}) is to be compared with its continuum counterpart.
The latter can be directly computed with the help of
the {\it self-reproducing} lattice theory \cite{Migd75,Witt91} defined via
the plaquette-factor (\ref{6.25m}) with
\be
e^{-F(R)}=dimR~exp[-f(\{C_{k}(R)\})]~.
\label{6.25xx}
\ee
Here $f(\{C_{k}(R)\})$ is some function of the Casimir operators
$C_{k}(R),~k=1,...,N,$ which parametrizes a particular $generalized$ $YM_{2}$
\cite{Witt91}. Note that the $q=1$ case of the family (\ref{4.1}) represents
the polynomial subvariety of (\ref{6.25xx}) for the particular choice of
$U=\hat{1}$.

The identification of $f(\{C_{k}(R)\})$, corresponding to a given
eigenvalue-system (\ref{6.25c}), would determine the universality class to
which the associated induced lattice gauge theory belongs. Instead of
addressing this question directly, we will find a lattice gauge theory
(\ref{6.25m}) which is supposed to have the same CL (in a finite physical
2-volume) as the model (\ref{6.25c}) in question. For this purpose,
in the infinite lattice-volume limit $L^{2},A\rightarrow{\infty}$ we
identify the pairs of the large $N$ gauge- and eigenvalue-systems for which
the large $A$ loop-averages are the same for generic values of the relevant
coupling constant(s) $\{b_{r}\}$.

To begin with, from the results of \cite{Rus,Witt91}, one readily deduces for
the disc-amplitude associated to (\ref{6.25m})
\be
\tilde{Z}_{G}({A}|U_{C})=
\sum_{R} (dimR)^{1-2G}~e^{-K(R){A}} 
~\chi_{R}(U_{C})~,
\label{8.11}
\ee
where $K(R)={F}(R)+ln(dimR),$. To make a crosscheck, one observes that for
the Heat-Kernal action (\ref{6.25m}) (with $f(\{C_{k}(R)\})\sim{C_{2}(R)}$)
eq. (\ref{8.11}) reproduces the canonical expression derived in
\cite{Rus,Witt91}.
Compared to the pattern (\ref{8.2}), the above formula contains the
topological piece $(dimR)^{1-2G}$ (which is $A$-independent) and depends on
the eigenvalues of the single holonomy $U_{C}$. As a result, the large $N$
partitition function $\tilde{X}^{(G)}_{L^{2}}$ on a discretized surface of
genus $G$ assumes the form
\be
\lim_{N\rightarrow{\infty}}\sum_{R} (dimR)^{2-2G}~e^{-K(R)L^{2}}=
[~\sum_{R} (dimR)^{\frac{2-2G}{2L^{2}}}~e^{-\frac{K(R)}{2}}~]^{2L^{2}}~,
\label{8.13}
\ee
where the weight $K(R)$ is supposed to be consistent with the
$O(N^{2})$-scaling of $ln[\tilde{X}^{(G)}_{L^{2}}]$. To be more specific,
in what follows we concentrate on the  case when
$e^{-F(R)}$ in eq. (\ref{6.25m}) is restricted to the family (\ref{4.1}).
For a preliminary determination of the pairing, one observes that the
pattern (\ref{8.13}) precisely matches with that of
eq. (\ref{8.5}) provided the identification 
\be
{F}(R)+ln(dimR)\equiv{K(R)}=S(R)~~,~~L^{2}\rightarrow{\infty}~~,~~\forall{G}~,
\label{8.14}
\ee
complemented by the infinite lattice-volume limit. To ensure
that the free energy (in both (\ref{6.25m}) and in the associated via
(\ref{8.14}) eigenvalue-system (\ref{6.25c})) is $\sim{O(N^{2})}$, one
is to require that the pattern (\ref{4.1}) of $e^{-F(R)}$ satisfies
$q>max[(1-({2-2G})/{2L^{2}})~;~1]$.
Note also that, in the $G=1$ case of the 2-tora, the matching (\ref{8.14})
accidentally holds true for any finite $L^{2}$ as well.

Next, consider an arbitrary nonself-intersecting loop $C$ of the area $A$ on a
generic discretized $2d$ surface of genus $G$ with the total area $L^{2}$.
For simplicity, we restrict our attention to the case when, after cutting
along the contour $C$, the disc of the area $A$ has $zero$ number of handles.
Employing \cite{Rus,Witt91}, one derives
\be
<W^{R_{0}}_{C}>=\frac{1}{\tilde{X}^{(G)}_{L^{2}}}\sum_{R}
(dimR)^{2-2G}~e^{-L^{2}K(R)}~
\tilde{B}(R|R_{0},A)~,
\label{8.15}
\ee
\be
\tilde{B}(R|R_{0},A)=\sum_{R_{1}}\frac{dimR_{1}}{dimR}~
e^{-[K(R_{1})-K(R)]A}~L_{R_{1}|R,R_{0}}~.
\label{8.16}
\ee
To simplify the comparison with (\ref{8.9}), in what follows $R_{0}$ is
selected to be the fundamental irrep $f$. According to the representation
theory \cite{Gr-in-phys}, it substantially reduces the pattern of the GLR
fusion rules
\be
\{L_{R_{1}|R,f}=1~or~0\}~\Longrightarrow
{~(L_{R_{1}|R,f})^{L_{C}}=L_{R_{1}|R,f}}~
\label{8.17}
\ee
which makes the relevant analysis particularly transparent.
Combining (\ref{8.17}) with the identification (\ref{8.14}), one observes
that the peculiar 'renormalization' of the perimeter-law
by the factor $(dim~f)^{1-L_{C}}=N^{1-L_{C}}$
is the only difference between (\ref{8.9}) and (\ref{8.15}) in the limit
$L^{2},A\rightarrow{\infty}$. More precisely, denote the
effective actions in the integral representation (akin to eq. (\ref{A.14v}))
for the loop-average
\be
<W^{f}_{C}>=N^{H_{k}(L_{C})}\int dh~e^{-B^{(k)}_{eff}(h|A,\{b_{r}\})}
\label{A.14b}
\ee
by $B^{(k)}_{eff}(h|A,\{b_{r}\})$ where $k=1,2$ stand for the options
(\ref{8.9}) and (\ref{8.15}) respectively, and $H_{1}(L)=2-L,~H_{2}(L)=1$.
According to the explicit form (\ref{4.1}) of $e^{-S(R)}=e^{-K(R)}$, the
large $A$ limit $\lim_{A\rightarrow{\infty}} S^{(k)}_{eff}(h|A,\{b_{r}\})$ is
$k$-independent for generic values of $\{b_{r}\}$. To be more specific,
consider the continuum limit $\tilde{b}^{2}\rightarrow{0}$ and take
the $M_{0}=1$ $U(N)$ model (\ref{4.1c}). In the limit
$\tilde{b}^{2}\rightarrow{\infty}$ the spectral density
(\ref{4.55x}) tends to zero: $\rho(\eta)\sim{O(\tilde{b})}$.
As it is clear from Appendix B, this scaling of $\rho(\eta)$ implies that
$[dimR_{1}/dimR]\rightarrow{1}+O(\tilde{b}^{2})$ (for admissible
$\eta^{2}<2q/\tilde{b}^{2}$ and the characteristic values of $R$). In other
words, the $[dimR_{1}/dimR]$-dependent mismatch between (\ref{8.10}) and
(\ref{8.16}) is expected to be irrelevant in the continuum limit.

Upon a reflection, the above correspondence between (\ref{8.9}) and
(\ref{8.15}) suggests the following equivalence. The gauge system (on a sphere of
a $finite$ physical 2-volume) induced from (\ref{6.25c}) and the associated
ordinary gauge theory (\ref{6.25m}) in $D=2$ have one and the
same continuum limit (CL) provided the identification (\ref{8.14}). To
justify the general consistency of this statement, we observe first that once the
localization (\ref{6.70}) takes place
in the $2d$ system (\ref{6.25c}) then similar localization (with the same
scaling) is valid in (\ref{6.25m}) associated via (\ref{8.14}). The converse
statement is also true. This
correspondence between the localizations follows from the specific
relation (\ref{8.14}) between the pair of the weights $e^{-F(R)}$ and
$e^{-S(R)}$. To make it manifest, one simply needs to change the variables
going over from $\lambda_{j}$- to the $\chi_{j}=([\tilde{g}N^{\frac{1}{2}}]
\lambda_{j}/N)$-fields like in eq. (\ref{4.21x}).

Next, the localization in the ordinary $D=2$ lattice gauge theory indeed
implies that the latter theory approaches the continuum. It is predetermined
by the absence of the propagating degrees of freedom in the $2d$ gauge
systems so that the renorm-group anomalous dimensions vanish.
To see how it works, introduce the infinitesimal lattice spacing $a$ and some
$finite$ parameter $\kappa \sim{\sqrt{\sigma}}$ responsible for the physical
mass-scale. Then, given the localization-pattern (\ref{7.2n}) of the lattice
action (\ref{7.2m}), the judiciously adjusted scaling
\be
A_{\rho}=B_{\rho}a~~,~~\tilde{b}={\kappa a}~~~;~~~
\lim_{a\rightarrow{0}}~\tilde{b}^{2}L^{2}=const~,
\label{7.2vv}
\ee
in the limit $a\rightarrow{0}$ is supposed to directly provide with the
pattern of the continuum gauge action associated to the corresponding
low-energy theory of (\ref{6.25m}).

Finally, the equivalence of the CL in (\ref{6.25c}) and (\ref{6.25m}) is
tantamount to the equivalence of the effective theories for the
low-frequency modes of $B_{\rho}$. The perimeter-type $L_{C}$-dependent
terms, among other things,
are responsible for the details of the $short$-distance attachment of
the colour-electric flux to the external Wilson loop source. As such, they
may be sensitive to the details of the discretization (of both the space-time
and the action) that in our case is
reflected by the mismatch in $H_{k}(L_{C})$ of eq. (\ref{A.14b}). On the
contrary, in the CL limit (\ref{7.2vv}) (keeping $A/L^{2}$ finite),
the bulk $(\tilde{b}^{2}L^{2}),(\tilde{b}^{2}A)$-dependent terms in both
(\ref{8.9}) and (\ref{8.15}) are supposed to be low-energy quatities
insensitive to the lattice discretization.

Formally, one could consider instead of a single
loop average the so-called Creutz ratios (see e.g. \cite{Creutz}) of the
products of the averages
\be
\frac{<W^{f}_{C_{1}}><W^{f}_{C_{2}}>}{<W^{f}_{C_{3}}><W^{f}_{C_{4}}>}~~~,~~~
L_{+}=L_{C_{1}}+L_{C_{2}}=L_{C_{3}}+L_{C_{4}},
\label{8.19}
\ee
where the pairs of the contours $\{C_{1},C_{2}\}$ and $\{C_{3},C_{4}\}$ are
constrained to have the same overall perimeter $L_{+}$. Being originally
designed to cancel exactly the short-distance perimeter-type contributions,
the ratios are evidently insensitive to the difference in $H_{k}(L_{C})$ of
eq. (\ref{A.14b}). Altogether, the above general arguments are consistent
with the presumable equivalence of the universality class to which the
latter pairs of the ordinary and the induced $2d$ gauge theories belong.
Note also that, in the case of a
generic irrep $R_{0}$, the mismatch between the patterns (\ref{8.9}) and
(\ref{8.15}) includes additionally the perimeter-type contribution owing to
the different powers of $L_{R_{1}|R,R_{0}}$. As it is clear from the above
discussion, this piece of the answer is irrelevant for the determination of
the universality class (even for a finite $N$) which is consistent with the
previous consideration based on the identification (\ref{8.14}).

\section{The large $N$ pattern of the loop-averages.}
\setcounter{equation}{0}

Building on the readily computable case of the $D=2$ eigenvalue-systems
(\ref{6.25c}), let us now discuss the general
pattern of the large $N$ scaling including self-intersecting contours and
multi-loop averages. We will argue in particular that, in the gauge theories
induced from (\ref{6.25c}), the standard large $N$ factorization
takes place. The scaling of a generic irreducible interaction between a
number of Wilson contours is also considered. In
conclusion, we comment on peculiar reasons behind the unconventional
$N\rightarrow{\infty}$ scaling (\ref{A.14v})
\be
\lim_{N\rightarrow{\infty}}<W^{f}_{C}>\sim{O(N^{2-L_{C}})}
\label{8.18b}
\ee
of the average of a single nonself-intersecting loop (\ref{2.13}).

It is instructive to begin with the pattern of the irreducible interactions
between a given number of nonintersecting, nonself-intersecting Wilson
loops $C_{p},~p=1,...,B,$ on a particular $2d$ discretized surface of the area
$L^{2}$. As usual, the corresponding amplitudes are defined by the irreducible
correlators of $B$th order
\be
<<\prod_{p=1}^{B}W^{f}_{C_{p}}>>=<\prod_{p=1}^{B}W^{f}_{C_{p}}>-
\prod_{p=1}^{B}<W^{f}_{C_{p}}>-...~~.
\label{8.25}
\ee
They are deduced from the ordinary $B$-loop correlators
$<\prod_{p=1}^{B}W^{f}_{C_{p}}>$ via subtraction of all lower order
reducible contributions as prescribed by the standard cluster expansion.
One may presume that the pattern (\ref{7.2m}),(\ref{7.2n}) 
is universal for a wide class of the lattice gauge theories (and in particular
for the family (\ref{6.25m}) with $e^{-F(R)}$ given by (\ref{4.1})) with the
free energy $\sim{O(N^{2})}$. Then,
provided the $absence$ of 'unnatural' cancellation between individual
diagrams, the 't Hooft double-line representation (for the large $N$ WC
series) yields the topological expansion in the form
\be
<<\prod_{p=1}^{B}W^{f}_{C_{p}}>>=\sum_{H\in{\bf Z_{\geq{0}}}}
w(B,H)~N^{2-2H-B}~,
\label{8.26}
\ee
where $w(B,H)\sim{O(N^{0})}$ and $H$ denotes the genus of the abstract surface
in the {\it index}-space. The same pattern is supposed to arise in the context
of the large $N$ strong coupling expansion on a lattice.

To deduce the large $N$ pattern of $<<\prod_{p=1}^{B}W^{f}_{C_{p}}>>$ in the
induced gauge theories, I propose to compare the $finite$ $N$ averages
$<\prod_{p=1}^{B}W^{f}_{C_{p}}>$ in the ordinary $2d$ gauge system
(\ref{6.25m}) and in the eigenvalue-model (\ref{6.25c}) associated via
(\ref{8.14}). To this aim, one is to employ the cutting-gluing
technique \cite{Migd75,Witt91} already used in the previous subsection.
For simplicity, we assume additionally that (nonself-intersecting) $C_{p}$
are not homotopic to the defining uncontractible cycles of the surface on
which both of the associated sytems are defined. In this case, 
the conventional pattern (\ref{8.26}) gets modified (as it will be clear in
the continuum limit) by the perimeter-type factors
\be
<<\prod_{p=1}^{B}W^{f}_{C_{p}}>>=(\prod_{p=1}^{B}N^{1-L_{C_{p}}})~
\sum_{H\in{\bf Z_{\geq{0}}}}
\tilde{w}(B,H)~N^{2-2H-B}~,
\label{8.29}
\ee
with $\tilde{w}(B,H)\sim{O(N^{0})}$.

To demonstrate (\ref{8.29}), we first cut the surface along the set
$\{C_{p}\}$ of the contours that results in a set of disjoint $2d$ windows.
Let the $q$th window have the area $A_{q}$ being endowed with certain number
$B_{q}$ of boundary discs and that $G_{q}$ of handles. Similarly to the
simplest $B=1$ case (\ref{8.2}), one evaluates the corresponding
$B_{q}$-disc amplitude
\be
Z^{B_{q}}_{G_{q}}({A}_{q}|\{U({\bf z^{(p)}_{k}})\})=
\sum_{R} e^{-S(R){A}_{q}}~ \prod_{p=1}^{B_{q}}
\prod_{{\bf z^{(p)}_{k}}\in{C_{p}}}
\chi_{R}({U}_{\rho_{k}}({\bf z^{(p)}_{k}}))~.
\label{8.27}
\ee
This expression is to be confronted with the corresponding $B$-disc amplitude
in the ordinary $2d$ gauge theory
\be
\tilde{Z}^{B_{q}}_{G_{q}}({A}_{q}|\{U_{C_{p}}\})=
\sum_{R} (dimR)^{2-2G_{q}-B_{q}}~e^{-K(R){A}_{q}} 
~\prod_{p=1}^{B_{q}}\chi_{R}(U_{C_{p}})~,
\label{8.28}
\ee
where $K(R)$ is defined by eq. (\ref{8.14}).

By the same token as in eq. (\ref{8.3}), the $B$-loop correlator
$<\prod_{p=1}^{B}W^{f}_{C_{p}}>$ is to be composed from the
associated amplitudes (\ref{8.27}) (or (\ref{8.28}) respectively) performing
the remaining Haar integrations over $\{U({\bf z^{(p)}_{k}})\}$. Apparently,
this procedure can be visualized then as gluing
the above windows back so that their boundaries sandwitch properly
the loops $W^{f}_{C_{p}}$ involved. Thus, the only difference between the
ordinary
$2d$ gauge theory and the associated $2d$ eigenvalue-system is in the
mismatch between the patterns of the multi-disc factors (\ref{8.27}) and
(\ref{8.28}). The most transparent
situation arises in the continuum limit (\ref{7.2vv}) when all ratios
$A_{q}/L^{2}$ are kept finite. A minor modification of the arguments (see Section 5.1)
suggests that, in this limit, the identification (\ref{8.14}) ensures the
proportionality of the $finite$ $N$ averages 
\be
<\prod_{p=1}^{B}W^{f}_{C_{p}}>_{eig.}\longrightarrow
{(\prod_{p=1}^{B}N^{1-L_{C_{p}}})~
<\prod_{p=1}^{B}W^{f}_{C_{p}}>_{gauge}}
\label{8.30}
\ee
evaluated in terms of the associated multi-disc amplitudes
(\ref{8.27}) and (\ref{8.28}) respectively. Presuming the 'smoothness' of the
continuum limit, one may expect that (\ref{8.29}) remains valid for 
generic values of $A_{q}$ and the coupling constants. In particular it
implies that in the eigenvalue-systems, despite the unconventional
perimeter-dependence (\ref{8.18b}), the large $N$ factorization of the
loop-correlators remains present.

Turning to the average associated to a single self-intersecting contour,
we first consider self-intersections on a finite number of {\it sites}. The
simplest example is the eight-figure
loop ('two-leaf flower') $C^{(2)}=C_{1}({\bf x_{0}})\cup C_{2}({\bf x_{0}})$
composed of the two nonself-intersecting contours $C_{k}({\bf x_{0}}),~k=1,2,$
which share the single site ${\bf x_{0}}$ in common. Similarly, one constructs
a contour $C^{(m)}=\cup_{k=1}^{m} C_{k}({\bf x_{0}})$ with the topology of
$m$-leaf flower based at ${\bf x_{0}}$. Employing the
eigenvalue-representation (\ref{8.7}), one observes
that for such contour the loop-average is $not$ an irreducible correlator
\be
\lim_{N\rightarrow{\infty}} <W^{f}_{\cup_{k=1}^{m} C_{k}({\bf x_{0}})}>
\rightarrow{N^{1-m}\prod_{k=1}^{m}<W^{f}_{C_{k}}>}
\sim{O(N^{m+1-\sum_{k=1}^{m}L_{C_{k}}})},
\label{8.21}
\ee
in contrast to the pattern of conventional gauge theories. 
Upon a reflection, the large $N$ scaling (\ref{8.21}) holds true for the
average of an arbitrary loop (\ref{2.13}) which after removing all the sites
with self-intersections is splitted into $m$ disjoint nonself-intersecting
contours $C_{k}$. Complementary, after subtraction of all the higher-order
'reducible' parts (contributing as
$\sim{O(N^{2+p-\sum_{k=1}^{m}L_{C_{k}}})},~p>0$)
the irreducible correlator (\ref{8.25}) of $m$th order is supposed to obey
(\ref{8.18b}). In particular,
for $m=2$ we expect that
\be
|<W^{f}_{\cup_{k=1}^{2} C_{k}({\bf x_{0}})}>-\frac{1}{N}
\prod_{k=1}^{2}<W^{f}_{C_{k}({\bf x_{0}})}>|
\sim{O(N^{2-L_{C_{1}}-L_{C_{2}}})}~.
\label{8.22}
\ee

In the case of a contour with self-intersections along links, the associated
average generically is not irreducible either. The additional reason here is
the existence of the (irreducible) one-link correlators (with
$\forall{R_{k}}\in{Y^{(N)}_{n_{k}}}$)
\be
<\chi_{R_{1}}(U_{\rho}({\bf z}))
\chi_{R_{2}}(U^{+}_{\rho}({\bf z}))>\sim{O(\prod_{k=1}^{2}dimR_{k})}~~,~~
n_{1}=n_{2} \bmod N,
\label{8.18c}
\ee
which depend nontrivially on the relevant coupling
constants $\{b_{r}\}$. As a result, the scaling (\ref{8.18b}) is generically
modified as well.

\subsection{Comments on the local large $N$ $[{\bf Z_{N}}]^{D}$-symmetry.}

Next, let us return to the interpretation of the unconventional pattern
(\ref{8.18b}) of a nonself-intersecting loop average in the gauge theories
induced from the eigenvalue-systems (\ref{6.25c}). From the viewpoint of the
generic pattern (\ref{7.2m}),(\ref{7.2n}) of the matrix theories with the
$O(N^{2})$ free energy, the scaling (\ref{8.18b}) indicates the additional
$N^{1-L_{C}}$-suppression. In the more
general case (\ref{8.29}), the corresponding irreducible average is
down in magnitude by the factor $(\prod_{p=1}^{B}N^{1-L_{C_{p}}})$.
One might expect that (in the WC phase) the latter is the result of a specific
$cancellation$ between different Feynman diagrams (each scaling individually
in compliance with the
standard pattern (\ref{8.26})). As this cancellation is $not$ manifest within
the formal WC series, we have to return to the specifics of the original
eigenvalue-models. As we will see, in the framework of the saddle-point (SP)
method, the extra $(\prod_{p=1}^{B}N^{1-L_{C_{p}}})$-factor can be traced
back to the otherwise unobservable local $[{\bf Z_{N}}]^{D}$-symmetry
(\ref{2.50m}) of the leading $O(N^{2})$ order of the effective action
$S_{eff}(\{\omega^{(0)}\})$ (see eq. (\ref{1.35b})) restricted to
the SP orbit $\{\omega^{(0)}\}$.

To this aim observe first that, were not the invariance (\ref{2.50m}),
in the average of the product (\ref{2.13b}) each trace could be substituted
in the large $N$ limit by its SP value
$\chi_{f}(\omega^{(0)}(\rho_{k}|{\bf z_{k}}))$ (modulo (\ref{2.9}) and
the ${\bf Z_{N}}$ gauge transformations (\ref{2.50b}) both invisible for a
closed loop $C$). Altogether, it would
result in the conventional
$O(N)$-scaling and $trivial$ perimeter-law pattern of the large $N$
loop-average (\ref{2.13}). It is the additional
averaging over the $[{\bf Z_{N}}]^{D}$-orbit (\ref{2.50m}) which
ensures that (in the eigenvalue-systems (\ref{6.25c}))
the leading $O(N)$ contribution to $<W^{f}_{C}>$ is exactly cancelled.
Indeed, assume for a moment that one would neglect completely
the subleading $O(N^{1-\beta}),~\beta\geq{0},$ orders of
$S_{eff}(\{\omega(\phi)\})$ in eq. (\ref{1.35b}). In this 'approximation',
the fluctuations (in the background of the SP orbit
$\omega^{(0)}(\rho_{k}|{\bf z_{k}})$ generated by (\ref{2.50b}), (\ref{2.9})
and (\ref{1.2x}))
induced by the infinitesimal transformations (\ref{2.50m}) become exactly
$zero$ modes. The contribution of these modes would render an arbitrary
correlator (\ref{2.13b}) identically zero for any
nonself-intersecting loop $C$
\be
<W^{f}_{C}>\rightarrow{N^{1-L_{C}}}\prod_{\{{\bf z_{k}}\in{L_{C}\}}}~
\sum_{t({\bf z_{k}})\in{\bf Z_{N}}} t({\bf z_{k}})~
\chi_{f}(\omega^{(0)}(\rho_{k}|{\bf z_{k}}))=0
\label{8.31}
\ee
because such $W^{f}_{C}$ does $not$ contain the singlet component
with respect to the $[{\bf Z_{N}}]^{D}$-averaging over
$t({\bf z_{k}})\in{\bf Z_{N}}$.

The inclusion of the $subleading$ orders
of $S_{eff}(\{\omega(\phi)\})$ breaks (\ref{2.50m}) down to (\ref{2.50b}),
(\ref{2.9}) which makes the above modes {\it quasizero}. In turn, it
justifies that the large $N$ loop-averages (associated to a nonself-intersecting
contour) are nonvanishing. On the other hand, the
cancellation (\ref{8.31}) indeed foreshadows the
$N^{1-L_{C}}$-suppressed scaling (\ref{8.18b}). To make it manifest, consider
the 2-loop average
\be
<|W^{f}_{C}|^{2}>\sim{O(N^{2})}~,
\label{8.24}
\ee
with $|W^{f}_{C}|^{2}$ being invariant under (\ref{2.50m}). As it is clear
from the SP method, the leading $O(N^{2})$-order of (\ref{8.24}) is
accumulated
(in contrast to the ordinary gauge theories) by the reducible part composed
of the $n_{1}=n_{2}=1$ correlators (\ref{8.18c}). This substantiates that
the $N\rightarrow{\infty}$ average (\ref{2.13}) is necessarily suppressed
compared to the conventional $B=1$ $O(N)$-pattern (\ref{8.26}).

\section{Conclusions.}

In this paper we have proposed to induce a class of $SU(N)$ lattice gauge
theories from the novel family of the vector-field eigenvalue-models
(\ref{6.25c}) (or their generalizations) employing the mapping (\ref{2.0z}).
The latter models, in addition to the
local $[U(N)]^{D}$ conjugation-invariance (\ref{2.50}), are endowed with
the ${\bf Z_{N}}$ gauge symmetry (\ref{2.50b}) and global $[{\bf Z_{N}}]^{D}$
invariance (\ref{2.9}). The crucial consequence of (\ref{2.50}) is that $both$
the action (\ref{6.25c}) $and$ the Wilson loop averages (\ref{2.13b}) depend
nontrivially only on the $eigenvalues$ $\omega(\rho_{k}|{\bf z_{k}})$ of the
relevant link-variables.

Consequently, both the partitition function $\tilde{X}_{L^{D}}$ (composed
of (\ref{6.3n})) and
the loop-averages (\ref{2.13b}) can be rewritten in terms of the GLR
coefficients (\ref{6.3a}) parametrized by $O(N)$ integers. In turn, 
the $1/N$ expansion of both the free energy and (\ref{2.13b}) can be
reformulated as the semiclassical series provided
that $-ln[\tilde{X}_{L^{D}}]\sim{O(N^{2})}$. In this perspective,
the eigenvalue-models (\ref{6.25c}) facilitate a master-field representation
of the associated induced gauge systems. Recall also that in any literally
continuum system the local conjugation-invariance (\ref{2.50}) is $not$
consistent (even at the classical level) with the presence of any $finite$
number of derivative-dependent terms in the action. Thus, from the viewpoint
of the continuum $YM$ theory, the invariance (\ref{2.50}) of (\ref{6.25c}) is
to be interpreted as the huge auxiliary symmetry introduced by the judicious
$discretization$ (regularization) of both the action and the space-time.

In the large $N$ limit, the partition function of (\ref{6.25c}) is reproduced
(via (\ref{2.10})) by the ($L^{D}$th power of the) reduced generating
functional (\ref{6.3p}) of the GLR coefficients which can be further
transformed into the 1-matrix
representation (\ref{1.2b}). To be even more specific, the 1-matrix
eigenvalue-model in (\ref{1.2b}) is selected in the simplest form
(\ref{4.1}) which makes accessible the complete large $N$
phase structure of the $D\geq{2}$ induced gauge theories. In particular, we
pay the special attention to the issue of the continuum limit (CL) in the
latter theories. The important place in this
analysis is played by the scaling-condition (\ref{7.1}), imposed on the
effective 1-matrix system (\ref{1.2b}). As it is demonstrated, this condition
predetermines that in the induced gauge system the localization
$\{U_{\rho}({\bf z})\rightarrow{\hat{1}}\}$ of the link-variables takes place
which is tantamount to CL. Also the phenomenon of
the auxiliary 'continuum' limits is discussed.

The computation of the large $N$ Wilson loop averages $<W_{C}^{f}>$ is
particularly simple in the $D=2$ case where the concise representation
(\ref{A.14v}) is derived for generic nonself-intersecting loop $C$ on an
arbitrary $2d$ surface. Building on the transparent relation with
$<W_{C}^{f}>$ evaluated in the ordinary $2d$ gauge theories, we reveal
the peculiar modification (\ref{8.29}) of the standard pattern (\ref{8.26})
of the $1/N$-topological expansion. However the physical mass-scale (like the
string tension in (\ref{A.14v})) is adjusted to scale as $O(N^{0})$, and
the large $N$ factorization is expected to remain present. Therefore, we
conclude that
the additional $N^{1-L_{C}}$-suppression (\ref{8.18b}) does $not$ prevent to
consider the induced gauge systems as the elligible large $N$ cousins of the
conventional $SU(3)$ lattice gauge theories. Moreover, it is reasonable to
expect that the $O(N^{0})$ physical string tension (if any) in the induced
lattice $YM$ system reproduces that in the associated continuum large $N$
$YM$ theory. It is supported, in particular, by the explicit $D=2$
derivations of Section 5.1.

Definitely, the most ambitious goal of our project is to find an approach
for computation of the $D=4$ large $N$ loop-averages in the proposed induced
systems adjusted to have the proper continuum limit. If successful, it 
might provide with the framework appropriate to address the issue
of confinement in the standard $D=4$ $continuum$ $SU(N)$ theory in the limit
$N\rightarrow{\infty}$.
Unfortunately, the $D=2$ technique of Section 5, being directly generalized to
$D\geq{3}$, does not seem to provide with a practical scheme. More promising
direction is to synthesize the latter technique with the $1/N$ saddle-point
method applied to the eigenvalues $\omega(\rho_{k}|{\bf z_{k}})$ employing
the appropriate representation (\ref{2.13b}) of
$\lim_{N\rightarrow{\infty}}<W_{C}^{f}>$.
The actual analysis in this framework is additionally complicated by the
presence of the quasizero modes associated to
the $[{\bf Z_{N}}]^{D}$ symmetry (\ref{2.50m}) of the leading $O(N^{2})$ order
of the effective action $S_{eff}(\{\omega^{(0)}\})$ (defined by eq.
(\ref{1.35b})) restricted to the SP orbit $\{\omega^{(0)}\}$. As
a result, even in the computation of $\lim_{N\rightarrow{\infty}}<W_{C}^{f}>$
it looks necessary to take into account the $subleading$ orders of
$S_{eff}(\{\omega\})$.
Nevertheless, one might expect that the presumable large $N$ factorization
(\ref{8.29}) foreshadows potential simplifications. Clearly, it calls for a
new piece of technology to be developed.

\begin{center}
{\bf Acknowledgements.}
\end{center}

I am grateful to Yu.Makeenko and A.Polyakov for useful discussions and
would like to thank A.Losev, A.Mironov, A.Morozov, G.Semenoff for interesting
comments and questions. This project was started when the author was the
NATO/NCSE Fellow
at University of British Columbia, and I would like to thank all the stuff and
especially Gordon Semenoff for hospitality. The paper is also partially
supported by the CDRF grant RP1-253.

\app{The local $[{Z_{N}}]^{D}$-invariance.}

In this appendix we discuss the local $[{Z_{N}}]^{D}$-symmetry (\ref{2.50m})
of the leading $O(N^{2})$ order of the effective action
$S_{eff}(\{\omega^{(0)}\})$ (defined in eq. (\ref{1.35b})) restricted to the
SP values $\{\omega^{(0)}\}$.
Owing to $S(D(D-1)/2)$ invariance (\ref{6.3pc}), it suffies to consider the
simpler version of the $SU(N)$ plaquette-factor (\ref{6.25c}) with the sum
running over the single species $R_{2}$
\be
Z(\{\omega(\phi)\})=\sum_{R_{2}} e^{-S(R_{2})}
\prod_{\{\mu\nu\}} \chi_{R_{2}}(\omega(\mu|{\bf x}))...
\chi_{\bar{R}_{2}}(\omega(\nu|{\bf x})),
\label{C.1}
\ee
where $S(R_{2})=S(\bar{R}_{2})$ and $\phi\in{{\bf z}\otimes \rho}$.

Our purpose is to demonstrate that the $SU(N)$ SP solution
$\omega^{(0)}(\phi)$ is supposed to be {\it unique} and
${\bf Z_{2}}$-selfdual (where $\chi_{R}(\omega(\phi))=
[\chi_{R}(\bar{\omega}(\phi))]^{+}$)
\be
\omega^{(0)}_{j}(\phi)=-\omega^{(0)}_{N-j+1}(\phi)
\equiv{\bar{\omega}^{(0)}_{j}(\phi)}~
\bmod S(N)\otimes [{\bf Z_{N}}]^{D}~,
\label{C.2}
\ee
modulo the Weyl $S(N)$ group (\ref{1.2x})
combined with the $[{\bf Z_{N}}]^{D}$ transformations (\ref{2.50m}).
To this aim, one first observes that 
the factor (\ref{C.1}) is by construction
${\bf Z_{2}}$ selfdual $Z(\{\omega(\phi)\})=Z^{+}(\{\omega(\phi)\})=
Z(\{\bar{\omega}(\phi)\})$. It implies that on the SP orbit, generated by the
relevant symmetries to be determined, there is
a point corresponding to a ${\bf Z_{2}}$ selfdual (modulo (\ref{1.2x}))
solution $\{\omega^{(0)}(\phi)\}$ .
Consider the SP {\it sub}orbit generated from the above ${\bf Z_{2}}$
invariant solution by the (local) Weyl group (\ref{1.2x}).
In this case, the corresponding stronger version of the
constraint (\ref{C.2}) is tantamount to the reality
of the character $\chi_{R}(\omega(\phi))=[\chi_{R}(\omega(\phi))]^{+}$
for $any$ (not necessarily selfdual) irrep $R$. In turn, the latter property
is sufficient for the selfconsistency of the localization (\ref{3.43b}) on
$R^{(0)}_{2}=\bar{R}^{(0)}_{2}$:
otherwise the effective action $S_{eff}(R_{2}|\{\omega(\phi)\})$ for $R_{2}$
\be
\sum_{R_{2}} e^{-S_{eff}(R_{2}|\{\omega(\phi)\})}=Z(\{\omega(\phi)\})
\label{C.3b}
\ee
would not be ${\bf Z_{2}}$-selfdual with respect to the interchange
$R_{2}\leftrightarrow{\bar{R}_{2}}$.

Now we are in a position to prove $[{\bf Z_{N}}]^{D}$ transformations
(\ref{2.50m}) of the above $S(N)$ suborbit $\{\omega^{(0)}(\phi)\}_{S(N)}$.
Let us introduce the twisted action
$S_{eff}(R_{2},t|\{\omega(\phi)\})$, resulting (akin to (\ref{C.3b}))
from eq. (\ref{C.1}) after the $[{\bf Z_{N}}]^{D}$-deformation
of the characters $\chi_{R}(V)$ involved,
\be
\chi_{R}(V)\rightarrow{\chi_{R}(tV)}=t^{n({R})}\chi_{R}(V)~~;~~
t=e^{i\frac{2\pi m}{N}}\in{Z_{N}}~,~R\in{Y_{n({R})}^{(N)}},
\label{C.4}
\ee
(where $n(R)$ is defined by (\ref{3.46m})) so that $Z(\{\omega(\phi)\})
\rightarrow{Z(\{\omega(\phi)\},t)}$. We claim that
$S_{eff}(R_{2},t|\{\omega^{(0)}(\phi)\})$ remains to
be ${\bf Z_{2}}$-selfdual
\be
S_{eff}(R_{2},t|\{\omega^{(0)}(\phi)\})=
S_{eff}(\bar{R}_{2},t|\{\omega^{(0)}(\phi)\})
\label{C.4x}
\ee
provided that $\{\omega^{(0)}(\phi)\}\in{\{\omega^{(0)}(\phi)\}_{S(N)}}$, i.e.
fulfils the stronger version of (\ref{C.2}) where the $[{\bf Z_{N}}]^{D}$
group is omitted. Moreover, we assert that (for such
$\{\omega^{(0)}(\phi)\}$) the resulting from (\ref{C.4}) sum (\ref{C.3b})
over $SU(N)$ irreps $R_{2}$ is localized
for $\forall{t\in{Z_{N}}}$ on the same $R^{(0)}_{2}(t)=R^{(0)}_{2}(1)
\equiv{R^{(0)}_{2}}$ as before $[{\bf Z_{N}}]^{D}$-twisting. In turn, it is
evidently tantamount to the invariance of the leading $O(N^{2})$ order of
$S_{eff}(\{\omega^{(0)}\})$ (defined by eq. (\ref{1.35b})) under
(\ref{2.50m}).

To prove these statements, one notes first that the explicit form
(\ref{4.6}) of ${\bf Z_{2}}$ transformation yields $exp(n(R)ln[t])=
exp(-n(\bar{R})ln[t])$. As only the real part of the overall twist-factor
contributes into $Z(\{\omega(\phi)\},t)$, the latter identity justifies
(\ref{C.4x}). In the
$SU(N)$ case (where $\lambda^{SU(N)}_{N}=0$ see (\ref{4.21})) it implies
that the saddle-point irrep is supposed to be selfdual $R^{(0)}_{2}(t)=
\bar{R}^{(0)}_{2}(t)$ irrespectively of $t$. To justify
$R^{(0)}_{2}(t)=R^{(0)}_{2}(1)$,
we employ the elligibility to restrict the large $N$ twisted sum (\ref{C.3b})
to that over $SU(N)$ selfdual irreps $R_{2}=\bar{R}_{2}$ for which
$N\lambda_{1}=2\sum_{i=1}^{N}\lambda_{i}\in{[2{\bf Z_{\geq{0}}}]}$ so that
${~t^{n({R}_{2})}=exp[i\pi (\lambda_{1}-N+1)]}$.
Finally, combination of the latter equation with the choice
$N\in{[2{\bf Z_{\geq{0}}}+1]}$  
results in $\lambda_{1}\in{[2{\bf Z_{\geq{0}}}]}$ so that $t^{n({R}_{2})}=1$.
Presuming the 'smoothness' of the large $N$ limit, it substantiates
$t$-independence of $R^{(0)}_{2}(t)$.

\app{The $D=2$ loop-averages.}

Consider the large $N$ $D=2$ average (\ref{8.9}) of a nonself-intersecting
Wilson
loop in the case when $R_{0}=f$ and the weight $e^{-S(R)}$ belongs to the
$U(N)$ subvariety (\ref{4.55vv}) of the pattern (\ref{4.1}). To begin
with, as we will see in a moment the factor $B_{L_{C}}(R|f,A)$ scales as
${O(N^{2-L_{C}})}$
provided the integers $\lambda_{j}$ (defining $R(\{\lambda\})$) are of order
of $\sim{O(N)}$ in the limit $N\rightarrow{\infty}$. We assert that 
the large $N$ sum in eq. (\ref{8.9}) is localized on the 
solution $R(\{\lambda^{(0)}\})\equiv{R(\bar{\lambda}^{(0)}(t))}$ of the SP
equations (\ref{4.21}). Indeed, recall first that a (unique, ${\bf
Z_{2}}$-selfdual) 
$R^{(0)}\equiv{R(\{\lambda^{(0)}\})}$ is the irrep on which
the large $N$ sum (\ref{8.5}) (defining the partitition function)
is localized. It is reasonable to presume that, for
$|\lambda^{(0)}_{j}-\lambda_{j}|\sim{O(\sqrt{N})}$, there is such $\alpha>0$
that
\be
\frac{|B_{L_{C}}(R(\{\lambda^{(0)}\})|f,A)-B_{L_{C}}(R(\{\lambda\})|f,A)|}
{B_{L_{C}}(R(\{\lambda^{(0)}\})|f,A)}\sim{O(N^{-\alpha})}~,
\label{A.2}
\ee
i.e. $B_{L_{C}}(R(\{\lambda\})|f,A)$ behaves smoothly on the
scale of the characteristic fluctuations of $\lambda_{j}$. Altogether, it
justifies the above assertion.

Next, one is to derive an explicit form of $B_{L_{C}}(R^{(0)}|f,A)$ in
terms of
the associated spectral density (\ref{4.55}). The representation theory
\cite{Gr-in-phys} tells that $L_{R_{1}|R^{(0)},f}$ is nonzero when the
associated to $R_{1}$ (double) Young table can be obtained from that
corresponding to $R^{(0)}$ by adding a single box in an admissible way.
The latter addition is elligble into the $i$th row if and only if the number
of boxes $n_{i}$ in this row (see eq. (\ref{3.46m})) is strictly less than in
the preceeding one
\be
n_{i}<n_{i-1}~\Longleftrightarrow{~(\lambda_{i-1}-\lambda_{i})}\geq{2}~.
\label{A.3}
\ee
According to the definitions (\ref{4.19}) and (\ref{4.55}) of
$d\bar{\lambda}(t)/dt$ and $\rho(\eta)$, it implies that in the domain of the
coupling constant(s) where $0\leq{\rho(\eta)}\leq{1/2},~\forall{\eta},$
the box can be generically added to $any$ row of the Young table
corresponding to $\rho(\eta)=
(dt/dh^{(0)}(t))|_{h^{(0)}(t)=\eta}$. Indeed, owing
to (\ref{A.2}) one can choose any discretization $\lambda^{(0)}_{j}/N,~
j=1,...,N,$ of the continuous function
$\bar{\lambda}^{(0)}(t)=h^{(0)}(t)+1/2$
which satisfies
$|N\bar{\lambda}^{(0)}(j/N)-\lambda^{(0)}_{j}|<<{\sqrt{N}}$. Condition
${\rho(\eta)}\leq{1/2}$ evidently ensures the existence of at least one such
discretization obeying (\ref{A.3}).

Due to (\ref{4.21x}), the constraint ${\rho(\eta)}\leq{1/2}$ is always
fulfilled in the large $N$ WC limit $\{b_{r}\rightarrow{0}\}$ relevant for the
analysis of the continuum limit. Therefore in what follows, we will restrict
our attention to the cases when ${\rho(\eta)}\leq{1/2}$ and thus (\ref{A.3}))
is valid. As a result, (by the same token as in \cite{Boulat}) one obtains 
\be
\sum_{R_{1}}e^{-[S(R_{1})-S(R^{(0)})]A}L_{R_{1}|R^{(0)},f}=
\int d\eta~ \rho(\eta)~[T(\eta)]^{{qA}}~
e^{-\frac{dV(\eta)}{d\eta}A},
\label{A.5}
\ee
\be
T(h(\frac{k}{N}))=exp[\sum^{N-1}_{i\neq{k}}
ln(1+\frac{1/N}{h(\frac{k}{N})-h(\frac{i}{N})})]~,
\label{A.5b}
\ee
where we have skipped the upperscript: $h^{(0)}(t)\equiv{h(t)}$. In the
derivation of the above equation, we employ that when the box is added in
$k$th row (of the Young table associated to $R^{(0)}$) then
\be
V(R_{1})-V(R^{(0)})\rightarrow{\sum_{m=1}^{M_{0}}}
[\tilde{b}_{2m}]^{2m}
\frac{[\lambda^{(0)}_{k}-\frac{N-1}{2}]^{2m-1}}{N^{2m-1}}
\rightarrow{\frac{dV(\eta)}{d\eta}|_{\eta=h(\frac{k}{N})}},
\label{A.6}
\ee
\be
T(h(\frac{k}{N}))=\left(\frac{dimR_{1}}{dimR^{(0)}}\right)=
\frac{\prod_{i\neq{k}}(\lambda^{(0)}_{k}+1-\lambda^{(0)}_{i})}
{\prod_{i\neq{k}}(\lambda^{(0)}_{k}-\lambda^{(0)}_{i})}
\label{A.7}
\ee
where $V(R(\{\lambda\}))$ is defined by eq. (\ref{4.55vv}).

The remaining step is to find the continuum representation for $T(\eta)$
in terms of $\rho(\eta)$. For this purpose, one is to expand the logarithm in
eq. (\ref{A.5b})
\be
T(h(\frac{k}{N}))=-(\sum^{|i-k|\leq{M}}_{i\neq{k}}+\sum_{|i-k|>{M}})
~\sum_{p\geq{1}}~
\frac{1}{p}[N({h(\frac{i}{N})-h(\frac{k}{N})})]^{-p}~,
\label{A.8}
\ee
treating $({h(\frac{k}{N})-h(\frac{i}{N})})$ differently for small,
$|i-k|\leq{M}$, and large, $|i-k|>{M}$, values of $|i-k|$ (where
$M=N^{\gamma},~0<\gamma<1$). In the former domain, one is to
substitute
\be
[N({h(\frac{k}{N})-h(\frac{i}{N})})]\rightarrow
{(k-i)\frac{dh(t)}{dt}|_{t=\frac{k}{N}}}=
(k-i)\rho^{-1}(\eta)|_{\eta=h(\frac{k}{N})}~,
\label{A.9}
\ee
while in the latter case the (partial) sum over $i$ is to be transformed into
\be
\sum\limits_{|i-k|>{M}}
[N({h(\frac{i}{N})-h(\frac{k}{N})})]^{-p}\rightarrow{N^{1-p}
\int_{|x-\frac{k}{N}|>\frac{M}{N}}\frac{dx}
{[{h(x)-h(\frac{k}{N})}]^{p}}}.
\label{A.10}
\ee
As it is clear from (\ref{A.10}), the situation is somewhat distinct for the
$p=1$ and the $p\geq{2}$ terms of the expansion (\ref{A.8}). In the $p\geq{2}$
case, the 'long-wavelength' term (\ref{A.10}) does $not$ contribute into the leading
$O(N^{0})$ order (of $ln[T(\eta)]$) we are interested in for eq.
(\ref{A.5}). According to
(\ref{A.9}), the remaining 'short-wavelength' $p\geq{2}$ terms are represented
by the $convergent$ sums which results (see \cite{Boulat}) in
\be
\lim_{N\rightarrow{\infty}} \sum^{|i-k|\leq{M}}_{i\neq{k}}
~\sum_{p\geq{2}}~\frac{[\rho(\eta)]^{-p}}{p}(i-k)^{-p}=-
ln\left(\frac{sin(\pi\rho(\eta))}{\pi\rho(\eta)}\right)~,
\label{A.11}
\ee
where $h(\frac{k}{N})=\eta$ and the contribution of odd $p=2m+1$ vanishes.
Complementary, in the $p=1$ case the 'short-wavelength' contribution goes to
zero while the large $N$ limit of the 'long-wavelength' part merges with
the principle value of the integral $P \int
\frac{d\phi \rho(\phi)}{\eta-\phi}$. Combining all the pieces together
and employing the SP equations (\ref{4.55w}), one
arrives at (\ref{A.14v}).

\end{document}